\documentclass[numberedappendix]{emulateapj}

\usepackage{amsmath,amssymb,bm}
\usepackage{graphics,graphicx}

\usepackage{hyperref}
\usepackage{comment}
\usepackage{xspace}
\usepackage{natbib}
\usepackage{multirow}
\usepackage[percent]{overpic}
\usepackage{color}

\newcommand{\beq}{\begin{equation}}
\newcommand{\eeq}{\end{equation}}

\newcommand{\bfk}{\bm k}

\newcommand{\bfx}{\bm x}
\newcommand{\bfq}{\bm q}
\newcommand{\bfp}{\bm p}

\newcommand{\bftheta}{\bm \theta}

\newcommand{\Mpc}{\mathrm{Mpc}}
\newcommand{\Gpc}{\mathrm{Gpc}}

\newcommand{\resp}[1]{\textcolor{black}{#1}}
\newcommand{\respp}[1]{\textcolor{black}{#1}}

\shorttitle{Accurate fitting formula of matter bispectrum}  % must be less than 44 characters
\shortauthors{Takahashi et al.}

\begin{document}

% TN added the preprint number from YITP
\begin{flushright}
        \quad \\
        \quad \\
        \quad \\
        YITP-19-103\\
\end{flushright}

\title{\resp{Fitting the nonlinear matter bispectrum by the Halofit approach}}

\author{Ryuichi Takahashi$^{1}$, Takahiro Nishimichi$^{2,6}$, Toshiya Namikawa$^{3}$, Atsushi Taruya$^{2,6}$, Issha Kayo$^{4}$, Ken Osato$^{5}$, 
Yosuke Kobayashi$^{6}$, Masato Shirasaki$^{7}$}
%\newauthor

\affil{
$^{1}$Faculty of Science and Technology, Hirosaki University, 3 Bunkyo-cho, Hirosaki, Aomori 036-8588, Japan\\
$^{2}$Center for Gravitational Physics, Yukawa Institute for Theoretical Physics, Kyoto University, Kyoto 606-8502, Japan\\
$^{3}$Department of Applied Mathematics and Theoretical Physics, University of Cambridge, Cambridge CB3 0WA, UK, \\
$^{4}$Department of Liberal Arts, Tokyo University of Technology, Ota-ku, Tokyo 144-8650, Japan, \\
$^{5}$Institut d'Astrophysique de Paris, Sorbonne Universit\'e, CNRS, UMR 7095, 98bis boulevard Arago, 75014 Paris, France, \\
$^{6}$Kavli Institute for the Physics and Mathematics of the Universe (WPI), The University of Tokyo Institutes for Advanced Study (UTIAS),\\
The University of Tokyo, 5-1-5 Kashiwanoha, Kashiwa-shi, Chiba, 277-8583, Japan, \\
$^{7}$National Astronomical Observatory of Japan, Mitaka, Tokyo, 181-8588, Japan
}

\begin{abstract}

We provide a \respp{new} fitting formula of the matter bispectrum
in the nonlinear regime calibrated by high-resolution cosmological $N$-body simulations of $41$ cold dark matter ($w$CDM, $w=$ constant) models around the \textit{Planck} 2015 best-fit parameters.
As the parameterization in our fitting function is similar to that in Halofit,
%for the nonlinear matter power spectrum, 
our fitting is named BiHalofit.
The simulation volume is sufficiently large ($> 10 \, \Gpc^3$) to cover
almost \textit{all} measurable triangle bispectrum configurations in the universe. 
\resp{The function is also calibrated using one-loop perturbation theory at large scales ($k<0.3 \, h \, {\rm Mpc}^{-1}$).}
Our formula reproduced the matter bispectrum to within $10 \, (15) \, \%$ accuracy in  the \textit{Planck} 2015 model at wavenumber $k< 3 \, (10) \, h \, \Mpc^{-1}$ and redshifts $z=0\text{--}3$.
%\resp{Our formula can fit the simulations with $2.7 \, (3.4) \, \%$ r.m.s. deviation for wavenumber $k< 3 \, (10) \, h \, \Mpc^{-1}$ at redshifts $z=0\text{--}3$ for the \textit{Planck} 2015 model.}
\resp{The other $40$ $w$CDM models obtained poorer fits, with accuracy approximating $20 \, \%$ at $k<3 \, h \, \Mpc^{-1}$ and $z=0\text{--}1.5$ (the deviation includes the $10 \, \%$-level sample variance of the simulations).}
%\tnrev{The formula} also takes into account (
We also provide a fitting formula that corrects the baryonic effects such as radiative cooling and active galactic nucleus feedback, using the latest hydrodynamical simulation IllustrisTNG. 
We demonstrate that our new formula more accurately predicts the weak-lensing bispectrum than the existing fitting formulas.
This formula will assist current and future weak-lensing surveys and cosmic microwave background lensing experiments.
\respp{Numerical codes of the formula are available, written in Python\footnote{\url{https://toshiyan.github.io/clpdoc/html/basic/basic.html\#module-basic.bispec}}, C and Fortran\footnote{\url{http://cosmo.phys.hirosaki-u.ac.jp/takahasi/codes_e.htm}}.}
\end{abstract}

\keywords{gravitational lensing: weak -- methods: numerical -- cosmology: theory -- large-scale structure of universe}

\section{Introduction} \label{sec:intro}

Observations of the cosmic microwave background (CMB) have revealed that the primordial density fluctuations are well described by a Gaussian field \citep{Planck2018nonG}.
The statistical property of a Gaussian field is fully described by the two-point correlation function or its Fourier transform, the power spectrum (PS).
However, at late times, the density fluctuations become non-Gaussian through small-scale gravitational evolution.
To fully characterize the statistical property of the non-Gaussian field and to access its cosmological information beyond the two-point (2pt) statistics, higher-order statistics are required.
The leading correction term of the commonly used PS is the bispectrum (BS), the Fourier transform of the three-point (3pt) correlation function.

\resp{The 3pt correlation function was first measured in the angular clustering of galaxies \citep{Peebles1975,Groth1977}.
Several groups later measured the 3pt statistics in redshift space using spectroscopic survey data  \citep[e.g.,][]{Jing1998,Scocci2001,Kayo2004}.
%Theoretical models have been developed using perturbation theory and $N$-body simulation \cite[e.g.,][]{Fry1984,Fry1993}].
The use of 3pt statistics breaks a degeneracy between the galaxy bias and cosmological parameters \citep[e.g.,][]{Fry1993,Matar1997,Nishimichi2007}.
From the recent analyses for the Baryon Oscillation Spectroscopic Survey data\footnote{\url{https://www.sdss.org/surveys/boss/}}, as a part of the Sloan Digital Sky Survey\footnote{\url{https://www.sdss.org}}, the baryon acoustic oscillation features were detected in the BS and 3pt correlation function \citep{Gilmarin2016,Slepian2017}. The 3pt statistics contain valuable information complementary to the 2pt statistics and helped to tighten the constraints on the angular diameter distance to galaxies and the redshift space distortion.}

\resp{Among various observables of large-scale structure,} weak lensing can map a projected density field through the coherent distortion of background galaxies \cite[e.g.,][]{BS2001}.
Current active weak-lensing surveys include the Subaru Hyper Suprime-Cam (HSC)\footnote{\url{https://hsc.mtk.nao.ac.jp/ssp/}},
the Dark Energy Survey (DES)\footnote{\url{https://www.darkenergysurvey.org}}, and the Kilo-Degree Survey (KiDS)\footnote{\url{http://kids.strw.leidenuniv.nl}}.
These surveys have placed strong constraints on the cosmological parameters such as the matter density $\Omega_{\rm m}$ and the amplitude of density fluctuations $\sigma_8$ from the cosmic-shear two-point function \cite[e.g.,][]{DES2018,KiDS2018,Hamana2019,Hikage2019}.
In the 2020s, ground- and space-based missions such as the Large Synoptic Survey Telescope (LSST)\footnote{\url{https://www.lsst.org/}}, Wide Field Infrared Survey Telescope (WFIRST)\footnote{\url{https://wfirst.gsfc.nasa.gov/}}, and \textit{Euclid}\footnote{\url{https://www.euclid-ec.org/}} will commence operations.

The weak-lensing BS contains additional information that complements the PS.
Because it arises from the non-Gaussian properties, the BS is more sensitive to smaller-scale and lower-redshift structures than the PS. 
A joint analysis of both the PS and BS spectra breaks parameter degeneracy and provides tighter constraints \citep[e.g.,][]{TJ2004,Kilbin2005,Sefusatti2006,Munshi2011,Kayo2013b,Byun2017,Gatti2019}.
The BS can be comparable to or more powerful than the PS \citep{Berge2010,SN2013,Coulton2019}.
Several groups have derived useful constraints from the three-point cosmic-shear statistics of real data \citep{BMvW2002,Jarvis2004,Sembo2011,Van2013,Fu2014,Simon2015}. 
\resp{The higher-order moments of weak-lensing convergence also contain the non-Gaussian information \citep[e.g.,][]{Petri2015}. The DES will set the observational constraint from a joint analysis of the second- and third-order moments \citep{Cahng2018,Gatti2019}.}

CMB lensing is another promising cosmological probe of the density fluctuations at higher redshifts ($z \simeq 1\text{--}3$) than cosmic shear \citep[e.g.,][]{LC2006}.  
Recent CMB experiments have measured the lensing signals from temperature and polarization fluctuations \citep{Planck2018lens}. 
The CMB lensing-potential PS provides rich cosmological information that complements the information in galaxy weak lensing \citep[e.g.,][]{Planck2018lens}.
The BS and higher-order spectra representing the non-Gaussian density fluctuations would be important in future CMB lensing observations. 
The non-Gaussianity slightly affects the lensing PS \citep{PL2016} as well as the CMB temperature and polarization power spectra \citep{LP2016,Marozzi2018}.
It also contaminates the CMB lensing reconstruction \citep{Bohm2016,Beck2018,Fabbian2019}.
On the other hand, the lensing BS can be measured as a useful signal in future CMB experiments \citep{Namikawa2016}.

Against this background, an accurate model of nonlinear BS is highly demanded.
A nonlinear model of the PS with a few percent accuracy up to $k=10 \, h \, \Mpc^{-1}$ is also required to meet the statistical accuracy requirements of forthcoming weak-lensing surveys\footnote{To our knowledge, the required accuracy of the BS model for current and forthcoming surveys has not been estimated.}  \citep{HT2005,Hearin2012}.
\cite{SC01} calibrated a fitting formula of BS in $N$-body simulations, which was later improved by \cite{GM12}.
However, the squeezed BS computed by these formulas is double (in the worst cases) that obtained in the latest numerical simulations \citep{Fu2014,Coulton2019,Namikawa2019,Munshi2019}.
%They also did not include the baryonic effects, which become significant at $k \gtrsim 1 \, h {\rm Mpc}^{-1}$.
In this paper, we construct an improved fitting formula of the nonlinear matter BS calibrated in high-resolution cosmological $N$-body simulations of $41$ $w$CDM models (where $w$CDM refers to cold dark matter and dark energy with a constant equation of state $w$).
\resp{Mainly, we aim to construct the formula for the \textit{Planck} 2015 $\Lambda$CDM model up to $k=30 \, h \, \Mpc^{-1}$ in the redshift range $z=0\text{--}10$, hoping that the formula has little dependence on cosmology.
%by taking the form of a mapping from the PS to the BS, as was done in previous fitting formulas.
The other $40$ $w$CDM models supplement the calibration at relatively low redshifts ($z=0\text{--}1.5$). This allows us to explicitly examine the cosmological model dependence, which was not thoroughly done previously. We also include the calibration from one-loop perturbation theory at $k<0.3 \, h \, \Mpc^{-1}$ in the $z=0\text{--}10$ range, because the simulations have large sample variance at the largest scales.}
To ensure an accurate calibration, we bin the simulation data and theoretical prediction into wavenumbers $(k_1,k_2,k_3)$. 
We also consider the baryonic effects in a public hydrodynamic simulation package called the IllustrisTNG suite \citep{Nelson2018}.

\resp{The outline of this paper is as follows. Section 2 discusses the basics of matter BS and gives the previous and our own fitting formulas. Section 3 details our simulations. Section 4 describes the fitting procedures and presents the resulting fitting function (Figures \ref{fig_bk_4configs}-\ref{fig_bk_allconfigs_residual_40models}). Section 5 discusses the baryonic effects on the BS using the IllustrisTNG data set. Section 6 compares the fitting formula predictions of the weak-lensing convergence BS with those of light-cone simulations. Section 7 discusses the systematics of cosmic-shear BS and CMB lensing BS. The main paper concludes with a summary in Section 8. Appendix A briefly discusses the halo model, and Appendixes B and C give the fitting formula and the baryonic correction, respectively.}

\section{Theory}

\subsection{Basics}

The cosmological density contrast is usually described by its Fourier transform $\tilde{\delta}(\bfk)$.
The matter PS and BS are respectively defined as
\begin{align}
 P(k_1) \, \delta_{\rm D}(\bfk_1+\bfk_2) &= \langle \tilde{\delta}(\bfk_1) \tilde{\delta}(\bfk_2) \rangle, \nonumber \\
 B(k_1,k_2,k_3) \, \delta_{\rm D}(\bfk_1+\bfk_2+\bfk_3) &= \langle \tilde{\delta}(\bfk_1) \tilde{\delta}(\bfk_2) \tilde{\delta}(\bfk_3) \rangle,
\end{align}
where $\delta_{\rm D}$ is the Dirac delta function.
Throughout this paper, we omit the redshift dependence in the arguments of functions because our discussion considers arbitrary redshifts. 
%the function arguments exclude the redshift because our discussion considers arbitrary redshifts.  

At the tree level (i.e., the leading order in perturbation theory), the matter BS is given by the product of the linear matter PS, $P_{\rm L}(k)$, as follows \citep[e.g.,][]{BCGS2002}:
\beq
 B_{\rm tree}(k_1,k_2,k_3) = 2 F_2(\bfk_1,\bfk_2) P_{\rm L}(k_1) P_{\rm L}(k_2) + 2 \, {\rm perm.}
\label{bk_tree}
\eeq
Here the last term describes two permutations $(\bfk_1,\bfk_2) \rightarrow (\bfk_2,\bfk_3)$ and $(\bfk_3,\bfk_1)$, which are applied to the wavevectors in the first term.
The $F_2$ kernel is
\beq
 F_2(\bfk_1,\bfk_2) = \frac{5}{7} + \frac{1}{2} \left( \frac{k_1}{k_2} + \frac{k_2}{k_1} \right) \mu_{12} + \frac{2}{7} \mu_{12}^2, 
\label{F2_kernel}
\eeq
where $\mu_{12}$ is the cosine of the angle between $\bfk_1$ and $\bfk_2$, i.e., $\mu_{12}=\bfk_1 \cdot \bfk_2 / (k_1 k_2)$.

To explore the nonlinear regime beyond the tree level, one usually relies on higher-order perturbation theories  \citep[e.g.,][]{Scocci1998,RW2012,Angulo2015,Hashimoto2017,Bose2018,LL2018}.
%The perturbative approaches are commonly used for the BS of galaxy clustering.
However, these are reliable only up to the mildly nonlinear regime ($k \lesssim 0.2 \, h \, \Mpc^{-1}$).
Another strategy adopts the analytical halo model \cite[e.g.,][]{CS2002}, which assumes that all matter is confined to halos.
This model is valid over a wide range of scales and redshifts, but its current accuracy is approximately $30 \%$  \cite[e.g.,][]{Lazanu2016,Bose2019}. 
The last one is a fitting function calibrated in $N$-body simulations over various scales, epochs and cosmological models.

\subsection{Previous fitting formulas}

\cite{SC01} (SC01) provided a fitting formula for the nonlinear BS.
Their function is similar to the tree-level formula (Eq.~\ref{bk_tree}), but replaces the linear PS with a nonlinear model and modifies the $F_2$ kernel to enhance the BS amplitude at small scales.
In the low-$k$ limit, their formula is consistent with the tree level.
In the high-$k$ limit, the BS is proportional to $P(k_1)P(k_2) + P(k_2)P(k_3) + P(k_3)P(k_1)$ according to the hyper-extended perturbation theory \citep{SF1999}. 
Their modified $F_2$ kernel contains six free parameters, which are fitted by their $N$-body results in four CDM models with $k<3 \, h \, \Mpc^{-1}$ and $0 \leq z \leq 1$.
Later, \citet[][hereafter GM12]{GM12} increased the number of free parameters in $F_2$ to nine and re-calibrated them from their $N$-body simulations in a single $\Lambda$CDM model over a relatively narrow range of wavenumbers ($k<0.4 \, h \, \Mpc^{-1}$) with $0 \leq z \leq 1.5$.
%Their fitting functions are
%\begin{align}
% B(k_1,k_2,k_3)
% = 2 F_2^{\rm GM/SC}(\bfk_1,\bfk_2) P(k_1) P(k_2) + 2 \, {\rm perm.}
%\end{align}

However, these formulas have several shortcomings.
First, they are based on a nonlinear PS model such as Halofit \citep{Smith2003,Takahasi2012}, HMcode \citep{Mead2015}, or \textsc{Cosmic Emulator} \citep{Lawrence2017}. 
This PS model needs to be prepared by the user along with the BS formula.
The discrepancies among these PS models are small but non-negligible, typically a few percent \citep[e.g.,][]{Schneider2016}; accordingly, they degrade the BS accuracy.
Second, as indicated by \cite{Namikawa2019}, these models overestimate the squeezed BS (i.e., the configuration of $k_1 \simeq k_2 \gg k_3$).
%One reason is that GM12 did not include the extreme squeezed case () 
Third, their fitting range of $k$ and $z$ is narrow. 
The current weak-lensing surveys measure the correlation function down to arcmin scales, requiring calibration up to $k=10 \, h \, \Mpc^{-1}$.
In addition, as the CMB lensing probes the high-redshift structures ($z \, \simeq 1\text{--}3)$, the calibration must extend at least to $z=3$. 
Finally, these models do not consider the baryonic effects, which are important at $k \gtrsim 1 \, h \, \Mpc^{-1}$.

\subsection{Our fitting formula}

Our fitting formula is based on the halo model and is similar to Halofit for the nonlinear PS \citep{Smith2003}.
The halo model is popular for evaluating the multi-point statistics of nonlinear density fields (the halo model BS is detailed in Appendix A).
%We construct a fitting formula based on this model.
Assuming that all particles are contained in halos, it decomposes the BS into three terms: one-, two- and three-halo terms (hereafter denoted as 1h, 2h and 3h, respectively).
The 1h term describes the correlation in an individual halo, and the 2h (3h) term accounts for the correlation among two (three) different halos.
The 1h and 3h terms dominate at small and large scales, respectively. 
Because the 2h term is subdominant in most of the triangle configurations \citep[except in the squeezed case; see, e.g.,][]{Valag2011,Valag2012}, it is dropped in our formulation and is absorbed by enhancing the 3h term at intermediate scales.

The fitting function consists of two terms,
\beq
  B(k_1,k_2,k_3) = B_{\rm 1h}(k_1,k_2,k_3) +  B_{\rm 3h}(k_1,k_2,k_3) ,
\label{bs_fit-func}
\eeq
and approaches the tree-level formula in the low-$k$ limit.
%We enhance the 3h term at intermediate scale to fill the gap.
The function contains $52$ free parameters to be fitted by our $N$-body data.
The fitting function is explicitly given in Appendix~B.

\resp{One may consider that $52$ free parameters are many. However, given the huge number of triangle configurations ($\sim 5 \times 10^5$) for all wavenumbers, redshifts, and cosmological models in our calibration, the number of parameters is rather small. In addition, recalling that %as 
the revised Halofit \citep{Takahasi2012} for the PS already contains $34$ free parameters, SC01 and GM12 contain $40$ and $43$ parameters in total, respectively (using the nonlinear PS from Halofit). Therefore, our parameters are only slightly more than in these previous models.}

\section{\resp{Numerical results}}

\begin{figure}
\vspace*{0.cm}
\begin{center}
\hspace*{-1.cm}
\includegraphics[width=11cm]{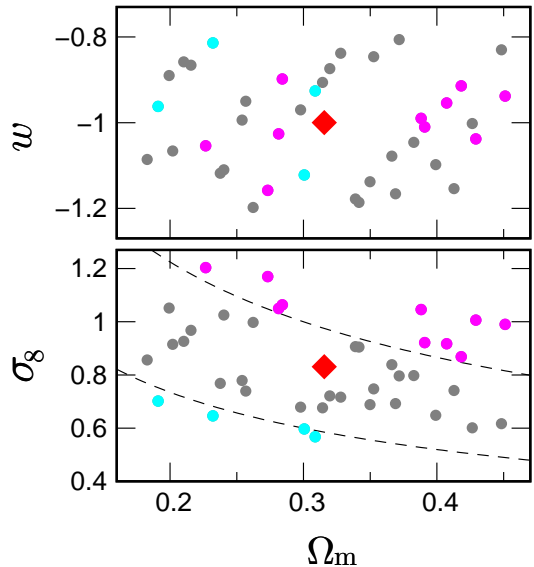}
\end{center}
\caption{Distribution of cosmological parameters in the $41$ models. The central red diamond is the \textit{Planck} 2015 best-fit $\Lambda$CDM, while the others are the $40$ $w$CDM models. The latter are divided into three groups of $S_8 \equiv \sigma_8 (\Omega_{\rm m}/0.3)^{0.5}$: $S_8>1.0$ (magenta circles), $S_8<0.6$ (cyan circles) and $0.6<S_8<1.0$ (gray circles). These three groups are separated by the boundaries (dashed lines) at $S_8=0.6$ and $1.0$.}
\vspace*{0.5cm}
\label{fig_cosmo_params}
\end{figure}

\begin{deluxetable*}{ccccccc}
\tablecaption{$N$-body Simulation Parameters}
\startdata
\hline 
 Cosmological & Box size & Number of & Number of & \resp{Particle Nyquist} & Maximum wavenumber & Output  \\
 model & ($h^{-1}\, \Gpc$) & particles & realizations & \resp{wavenumber ($h \, \Mpc^{-1}$)} & ($h \, \Mpc^{-1}$) & redshifts \\
  \hline
 \textit{Planck} 2015 $\Lambda$CDM & $4$   & $4096^3$ & $8$  & $3.22$ & $2.85$ & $0,0.55,1.03,1.48$ \& HighZ  \\
 & $2$   & $2048^3$ & $15$  & $3.22$ & $1.42$  & LowZ \\
 & $1$   & $2048^3$ & $21$  & $6.43$ & $28.5$  & LowZ \& HighZ \\
 & $0.2$ & $2048^3$  & $10$  & $32.2$ & $14.2$ & HighZ \\
   \hline
 $40$ $w$CDM & $2$   & $2048^3$ & $1$  & $3.22$ & $1.42$  &  LowZ  \\
 & $1$   & $2048^3$ & $1$ & $6.43$ & $28.5$  & LowZ 
\enddata
\tablecomments{
 In the output redshifts column, LowZ covers ten low redshifts ($z=$ $0, 0.15, 0.31, 0.42, 0.55, 0.69,$ $0.85, 1.03, 1.23,$ and $1.48$), and HighZ covers four high redshifts ($z=2,3,5,$ and $10$).
 The LowZ simulations with $L=1$ and $2 \, h^{-1}\, \Gpc$ are taken from \cite{Nishimichi2018}; the others are newly prepared in this work. 
}
\label{table_nbody}
%\vspace*{1.0cm}
\end{deluxetable*}

The fitting formula was calibrated in cosmological dark matter $N$-body simulations.
We used the $N$-body data set prepared by the \textsc{Dark Emulator} project \citep[][hereafter referred to as N19]{Nishimichi2018}.
%Baryons will be discussed in later section.
The N19 project has prepared $101$ flat cosmological models (a fiducial $\Lambda$CDM and additional $100$ $w$CDM models) in the range $z=0\text{--}1.48$.
%In this paper, we used $41$ models among them.
The project aims to emulate several halo observables such as the halo-matter correlation function, the halo mass function, and the halo bias for ongoing weak-lensing surveys.
The emulator will be publicly available soon.

%In this paper, we also run some additional $N$-body simulations to explore high $z$ (\geq 2) and high $k$.

\subsection{Cosmological models}

We used the N19 simulations of $41$ flat cosmological models\footnote{Unfortunately, the particle position data were lost for the rest (60) of the models owing to hard-disk trouble.}.
The fiducial $\Lambda$CDM model is consistent with the \textit{Planck} 2015 best fit \citep{Planck2015}, with matter density $\Omega_{\rm m}=1-\Omega_{\Lambda}=0.3156$, baryon density $\Omega_{\rm b}=0.0492$, Hubble \resp{parameter} $h=0.6727$, spectral index $n_{\rm s}=0.9645$, and amplitude of matter density fluctuation on the scale of $8 \, h^{-1} \, \Mpc$ $\sigma_8=0.831$.

The other $w$CDM models have six cosmological parameters: $\Omega_{\rm b}h^2, \Omega_{\rm cdm}h^2, \Omega_w, A_{\rm s}, n_{\rm s}$ and $w$.
Here the dark energy equation of state $w$ is assumed to be constant, and $A_{\rm s}$ is the amplitude of the primordial PS.
These parameters are distributed around the fiducial model in the ranges $\pm 5 \%$ for $\Omega_{\rm b} h^2$ and $n_{\rm s}$, $\pm 10\%$ for $\Omega_{\rm cdm} h^2$, and $\pm 20\%$ for $\Omega_w, \ln A_{\rm s}$ and $w$.
In the N19 project, the cosmological parameters were sampled using a Latin Hypercube Design \citep[e.g.,][]{Heitmann2009}.
The models were placed into five subsets, each containing $20$ models.
Figure~\ref{fig_cosmo_params} shows the distributions of $w$ and $\sigma_8$ vs. $\Omega_{\rm m}$ in the $41$ models (the fiducial $\Lambda$CDM models and two subsets of N19) considered in the present study.
The parameter range is wide enough for current and future weak-lensing surveys.
\resp{In fact, the current constraint from the HSC (DES) cosmic-shear 2pt statistics alone is $S_8=0.795^{+0.043}_{-0.047} ~(0.789^{+0.036}_{-0.038})$ in the flat $w$CDM model \citep{Hamana2019,Troxel2018}.}

Although these simulations are dark-matter-only simulations, their initial condition accounts for the free-streaming damping by massive neutrinos.
\resp{To compute the linear matter transfer function at the initial redshift of the simulations, N19 first computed the one at $z=0$ with massive neutrinos %(at $z=0$) by the square 
and then multiplied it by the ratio of the linear growth factor between $z=0$ and the target redshift, in which %without the neutrinos.
the scale-dependent growth due to neutrinos was neglected.
The same procedure was done in this work.
}
%This transfer function is used to calculate the initial PS.
The neutrino density in all models was fixed at $\Omega_\nu h^2=6.4 \times 10^{-4}$, corresponding to a total mass $0.06 \, \mathrm{eV}$. 
This $\Omega_\nu$ is included in $\Omega_{\rm m}$.

\subsection{$N$-body simulations}

Our simulation settings are summarized in Table~\ref{table_nbody}. To cover a wide range of length scales, we set four box sizes ($L=4,2,1,$ and $0.2 \, h^{-1} \, \Gpc$, where $L$ is the side length of the cubic box in the comoving scale). 
Note that the large simulation volume can include almost all measurable triangle configurations of BS in the real universe.
The large-volume simulations ($L= 4 \, h^{-1} \, \Gpc$) reduce the sample variance in the measured BS at small $k$, whereas the small-volume simulations ($L=0.2 \, h^{-1} \, \Gpc$) reveal the asymptotic behavior at high $z$.
Here the simulations with $L= 1$ and $2 \, h^{-1} \, \Gpc$ at $z=0 \text{--} 1.48$ are taken from N19, while the others are newly prepared in this work.
The largest- and smallest-box simulations supplement the dynamic range covered by N19.
The number of particles was set to $2048^3$ except for $L=4 \, h^{-1} \, \Gpc$ (where it was $4096^3$).
\resp{The resulting particle Nyquist wavenumber is $k_{\rm Ny}=\pi/l_{\rm p}$, where $l_{\rm p} \, (= n_{\rm p}^{-1/3})$ is the mean inter-particle separation at particle number density $n_{\rm p}$. The $k_{\rm Ny}$ values are listed in Table \ref{table_nbody}.}
The fiducial $\Lambda$CDM model has dozens of independent realizations, whereas each $w$CDM model has a single realization.

The initial matter PS was prepared by the public Boltzmann code \textsc{CAMB} \citep{CAMB2000}.
The initial particle distribution was determined by the second-order Lagrangian perturbation theory \citep[2LPT;][]{CPS2006,Nishimichi2009}\footnote{The 2LPT reduces the error in the BS estimate caused by transients from the initial condition to below $2 \, \%$ at $z \leq 1$ \citep{McCullagh2016}.} at redshifts of $z_{\rm in}=31,29,59,$ and $99$ for $L=4,2,1,$ and $0.2 \, h^{-1} \, \Gpc$, respectively.
The initial redshifts in the $40$ $w$CDM models were changed because the initial amplitudes differed among the models.
\resp{The initial redshift was determined by requiring the root-mean-square (rms) displacement to be $25 \%$ of the mean inter-particle separation to achieve the optimal balance between the artificial force due to the grid pre-initial configuration and the transient due to the truncation of the LPT at the second order.}
%Our simulations cover $z=0-10$.
The nonlinear gravitational evolution was followed using \resp{a tree-PM (particle mesh) code} \textsc{Gadget2} \citep{Springel2001,Springel2005}.
\resp{The number of PM grid cells was $4096^3$ ($8192^3$ for $L=4 \, h^{-1}\,\mathrm{Gpc}$). The gravitational softening length was set to $5 \%$ of the mean inter-particle separation. The \textsc{Gadget2} parameters (such as time step and force calculation parameters) were fine-tuned to determine the matter PS with percentage-level accuracy in N19 (see subsection 3.4 of their paper).}
The particle snapshots were dumped at $14$ redshifts ranging from $z=0$ to $10$ (see Table~\ref{table_nbody} for the exact redshift outputs).

To measure the density contrast, we assigned the $N$-body particles to the $1024^3$ regular grid cells in the box using the cloud-in-cell (CIC) interpolation with the interlacing scheme \citep[e.g.,][]{Jing2005,Sefusatti2016}.
The Fourier transform $\tilde{\delta}(\bfk)$ of the density field was then obtained by fast Fourier transform (FFT)\footnote{\textsc{FFTW} (Fast Fourier Transform in the West) is available at \url{http://www.fftw.org}.}.
To explore smaller scales, we also employed the folding method \citep{Jenkins1998}, 
\resp{which folds the particle positions $\bfx$ into a smaller box of side length $L/n$ by replacing $\bfx$ with $\bfx\%(L/n)$ (where $a\%b$ obtains a reminder of $a/b$).
This procedure effectively increases the resolution by $n$ times.
Here we set $n=4$ and $10$ at $L=4$ and $1 \, h^{-1} \, \Gpc$, respectively.}
%The wavenumber has discrete number, $\bfk=(2 \pi/L) \bfn$ where the components of $\bfn$ are integers. 
The minimum and maximum wavenumbers in the $1024^3$ cells were $k_{\rm min}^L= 2 \pi/L = 6.3 \times 10^{-3} \, h \, \Mpc^{-1} \, [L/(h^{-1} \, \Gpc)]^{-1}$ and $k_{\rm max}^L=512 \, k_{\rm min}^L = 3.2 \, h \, \Mpc^{-1} \, [L/(h^{-1} \, \Gpc)]^{-1}$, respectively. The folding scheme simply enlarged both $k_{\rm min}^L$ and $k_{\rm max}^L$ by $4$ or $10$ times. 
The resultant $k_{\rm max}^L$ values are given in Table~\ref{table_nbody}.

\subsection{Power spectrum measurement for accuracy check}

\begin{figure}
\vspace*{0.cm}
\begin{center}
\includegraphics[width=8.5cm]{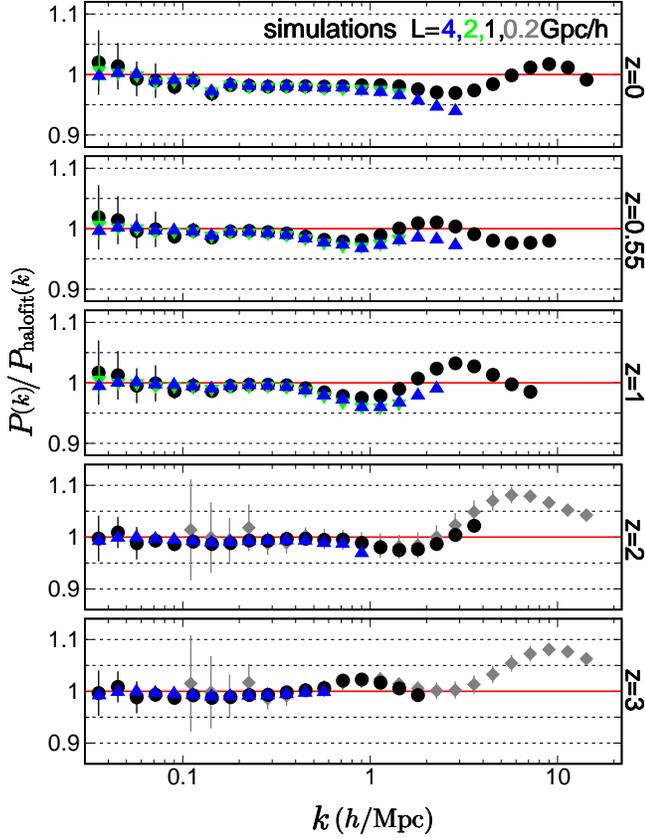}
\end{center}
\vspace*{-0.3cm}
\caption{Power spectrum ratio of the simulation to the Halofit prediction in the \textit{Planck} 2015 best-fit $\Lambda$CDM model. The results in simulation box sizes of $L=4,2,1,$ and $0.2 \, h^{-1} \, \Gpc$ are indicated by blue, green, black, and gray symbols, respectively. The shot-noise contribution is less than $3 \, \%$.}
\vspace*{0.cm}
\label{fig_pk_compare}
\end{figure}

The numerical accuracy was checked by comparing the simulated matter PS with the results of a previous fitting formula.
The PS estimator is given by
\beq
 \hat{P}(k) = \frac{1}{N_{\rm mode}^{\rm PS}} \sum_{|\bfk^\prime| \in k} \left| \tilde{\delta}(\bfk^\prime) \right|^2,
\eeq
where the summation is performed over $k-\Delta k/2<|\bfk^\prime|<k+\Delta k/2$ and $N_{\rm mode}^{\rm PS}$ is the number of modes in a fixed bin width ($\Delta \log_{10} k=0.1$).
Figure~\ref{fig_pk_compare} plots the PS ratio of the simulation to the revised Halofit prediction \citep{Smith2003,Takahasi2012}. 
We here plot the average $P(k)$ and its $1 \sigma$ error measured from the realizations.
The results in the different boxes were nicely consistent.
In larger simulation boxes, the measured PS was smaller than the Halofit prediction at large $k$ because of the lack of spatial resolution.
The shot noise was not subtracted because it contributed less than $3\,\%$ on the scales shown in Figure~\ref{fig_pk_compare}.
%Here, we do not subtract the shot noise, but instead show the $k$-range in which its contribution is less than $3 \, \%$. 
The simulations agreed with the fitting formula within $5 \, \%$ for $k<1 \, h \, \Mpc^{-1}$ at $z=0\text{--}10$ and $k<10 \, h \, \Mpc^{-1}$ at $z=0\text{--}1.5$.

\subsection{Bispectrum measurement}

The BS estimator is given by
\begin{align}
 \hat{B}(k_1,k_2,k_3) = \frac{1}{N_{\rm triangle}} \sum_{|\bfk_1^\prime| \in k_1} \sum_{|\bfk_2^\prime| \in k_2} \sum_{|\bfk_3^\prime| \in k_3} \nonumber \\
 \times \, \tilde{\delta}(\bfk_1^\prime) \tilde{\delta}(\bfk_2^\prime) \tilde{\delta}(\bfk_3^\prime) \delta^{\rm K}_{\bfk_1^\prime+\bfk_2^\prime+\bfk_3^\prime}
\label{bk_estimator}
\end{align}
where the summation is performed over all modes in the bin, $|\bfk_{\rm i}^\prime| \in k_{\rm i}$ (${\rm i}=1,2,3$), $N_{\rm triangle}$ is the number of triangles, and $\delta^{\rm K}$ is the Kronecker delta.
Throughout this paper, the log-scale bin width is constant ($\Delta \log_{10} k = 0.1$), \resp{unless otherwise stated}.
Equation \eqref{bk_estimator} was calculated by the FFT-based quick estimator \cite[e.g.,][]{Scocci2015}.
Using the identity $\delta^{\rm K}_{\bfk_1^\prime+\bfk_2^\prime+\bfk_3^\prime}=N_{\rm cell}^{-1} \sum_{\bfx} e^{ i ( \bfk_1^\prime+\bfk_2^\prime+\bfk_3^\prime) \cdot \bfx }$,
Eq.~\eqref{bk_estimator} reduces to
\begin{align}
  \hat{B}(k_1,k_2,k_3) = \frac{1}{N_{\rm triangle}} \frac{1}{N_{\rm  cell}} \sum_{
  \bfx} \left[ \sum_{|\bfk_1^\prime| \in k_1} \!\! \tilde{\delta}(\bfk_1^\prime) \, {\rm e}^{{\rm i} \bfk_1^\prime \cdot \bfx}
  \right. \nonumber \\
  \times \left. \sum_{|\bfk_2^\prime| \in k_2} \!\! \tilde{\delta}(\bfk_2^\prime) \, {\rm e}^{{\rm i} \bfk_2^\prime \cdot \bfx}
  \sum_{|\bfk_3^\prime| \in k_3} \!\! \tilde{\delta}(\bfk_3^\prime) \, {\rm e}^{{\rm i} \bfk_3^\prime \cdot \bfx}
  \right], \label{bk_estimator2}
\end{align}
where $\bfx$ is a discrete grid coordinate and $N_{\rm cell}=1024^3$ is the total number of cells.
The summation over $\bfk_{\rm i}^\prime$ is easily performed by FFT.
\resp{Although Eq.~(\ref{bk_estimator2}) can be quickly computed, the FFTs in all $k_{\rm i}$ bins require large memory resources. This demand limits the grid resolution ($N_{\rm cell}$).}

The shot noise is measured as 
\beq
\hat{B}_{\rm sn}(k_1,k_2,k_3) = \frac{1}{n_{\rm p}} \left[ \hat{P}(k_1) + \hat{P}(k_2) + \hat{P}(k_3) \right] - \frac{2} {n_{\rm p}^2}, \nonumber
\eeq
where $n_{\rm p}$ is the particle number density and $\hat{P}(k)$ is the PS estimator including the shot noise.

In the fiducial model, we calculated the average and standard deviation of BS from the realizations (the number of realizations is listed in Table~\ref{table_nbody}).
%We do not subtract the shot noise, instead we do not include the range where the shot noise is not negligible.
However, the results of the $w$CDM models have relatively large scatters because each $w$CDM model has only a single realization. 
Therefore, the fitting formula was the main calibration formula for the \textit{Planck} 2015 model, while the other models supplementarily checked its dependence on the cosmological parameters.

\section{Fitting procedure and results}

\subsection{Fitting to the $N$-body results}

\begin{figure}
\vspace*{0.cm}
\begin{center}
%\hspace*{-0.7cm}
\includegraphics[width=12cm]{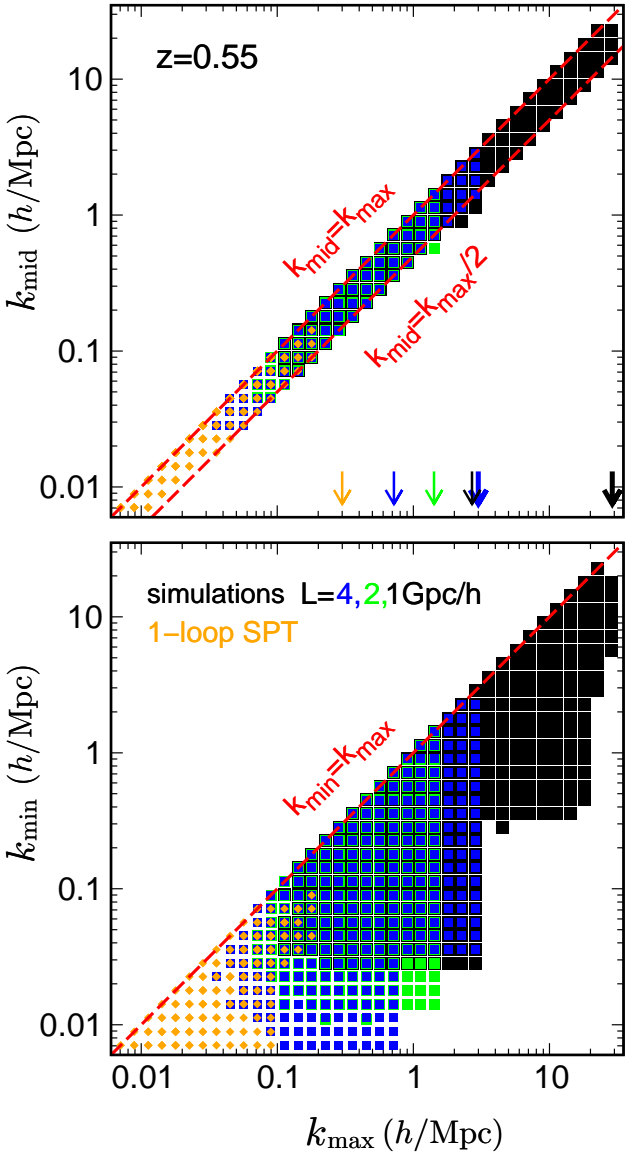}
\end{center}
\vspace*{-0.8cm}
\caption{Triangle configurations included in the calibration from the simulations with $L=4,2,1 \, h^{-1} \, \Gpc$ (blue, green and black squares) and from the one-loop standard perturbation theory (SPT; orange diamonds). Here $k_{\rm min},k_{\rm mid}$ and $k_{\rm max}$ are the minimum, middle and maximum side lengths of a triangle.
The dashed red lines correspond to particular triangles: $k_{\rm mid}=k_{\rm max}$ is the squeezed, $k_{\rm mid}=k_{\rm max}/2$ is the flattened, and $k_{\rm min}=k_{\rm max}$ is the equilateral. 
The arrows in the top panel indicate the maximum wavenumbers in the calibration from the simulations (blue, green and black) and from the SPT (orange). The thick (thin) arrows are with (without) the folding scheme.
All the points satisfy the conditions a) -- c) in subsection~4.1.}
\vspace{0.cm}
\label{fig_k1k2k3_region}
\end{figure}

\begin{figure}
\begin{center}
\includegraphics[width=9cm]{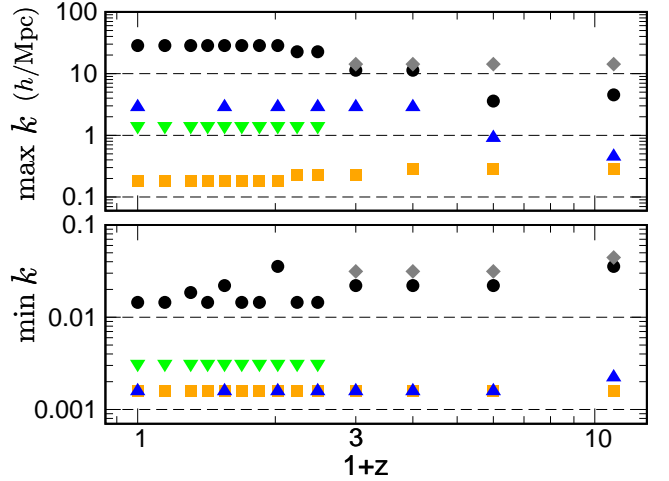}
\end{center}
\vspace*{-0.3cm}
\caption{\resp{Maximum and minimum wavenumbers included in the calibration as a function of $1+z$ (in log scale). Symbols denote the simulations with $L=4,2,1,$ and $0.2 \, h^{-1} \, \Mpc$ (blue, green, black and gray) and the one-loop SPT (orange).}
}
\label{fig_kmin_kmax}
\end{figure}

This subsection presents our fitting procedure to the $N$-body results.
The BS fitting function in Eq.~\eqref{bs_fit-func} contains $52$ parameters.
Arraying these parameters as $\bfp=(p_1,p_2,...)$, the best fit is determined by the standard chi-squared analysis:
\begin{align}
  \chi_{\rm sim}^2(\bfp) &= \resp{\sum_{c=1}^{41}  \sum_{z=0}^{10} \sum_{L} \sum_{k_1,k_2,k_3}^{30 \, h \, \mathrm{Mpc}^{-1}}} W_{c} W_z W_k \nonumber \\
   \times & \left[ \frac{B^{\rm bin}(k_1,k_2,k_3;\bfp) - B_{\rm sim}(k_1,k_2,k_3)}{\Delta B_{\rm sim}(k_1,k_2,k_3) + \epsilon(k_1,k_2,k_3)} \right]^2,
\label{chi_sim}
\end{align}
where the summation is performed for the $41$ cosmological models (subscripted by $c$), all redshifts $z=0\text{--}10$ (subscripted by $z$), all simulations in different box sizes $L$, and all triangles ($k_1,k_2,k_3$) up to $30 \, h \, \Mpc^{-1}$.
Here $B^{\rm bin}$ is the binned prediction of the fitting formula (given by Eq.~\ref{binned_BS_fit}), $B_{\rm sim}$ is the simulation result, and $\Delta B_{\rm sim}$ is the standard deviation estimated from the $N$-body realizations.
As each of the $40$ $w$CDM models has one realization \resp{and these cosmological parameters are basically similar\footnote{\resp{The differences of cosmological parameters among the $41$ models are less than $20 \, \%$ (see subsection 3.1).}}}, their relative standard deviations $\Delta \! \ln \! B_{\rm sim}$ $(\equiv \Delta B_{\rm sim}/B_{\rm sim})$ are assumed equal to those of the \textit{Planck} 2015.
\resp{Although a small change in $B_{\rm sim}$ may give a large impact on the resultant fitting formula, a change in $\Delta \! B_{\rm sim}$ would not significantly change the result.}
%As the error in the calibration (\ref{chi_sim}) is determined by the two terms ($\Delta \! B_{\rm sim}$ and $\epsilon$), the cosmological-model dependence of $\Delta \! B_{\rm sim}$ }
We also include a ``softening" term $\epsilon=0.02 \times B_{\rm sim}$ that reduces the influence of data points with very small \resp{$\Delta \! \ln \! B_{\rm sim}$ ($\ll \epsilon/B_{\rm sim} = 2 \, \%$)} \footnote{Since the number of realizations are not large enough for estimating the variance accurately, some data points accidentally have very small $\Delta B_{\rm sim}$.}.
\resp{At large (small) scales where $\Delta \! \ln \! B_{\rm sim}$ is larger (smaller) than $2 \, \%$, the $\Delta B_{\rm sim}$ ($\epsilon$) term dominates the denominator of Eq.~(\ref{chi_sim}). 
%At small scales ($\Delta \! \ln \! B_{\rm sim} < \epsilon/B_{\rm sim} (=2 \%)$), a small difference in $\Delta \! B_{\rm sim}$ would not significantly change the result.
The $\Delta B_{\rm sim}$ term gives more weight to smaller-scale data (because $\Delta \! \ln \! B_{\rm sim}$ is smaller), whereas the $\epsilon$ term gives an equal weight, irrespective of scale.}
The weight factors ($W$) were introduced to place greater importance on the lower-redshift data ($W_z$) because cosmic shear probes the low-$z$ ($\lesssim 0.5$) structures, and on larger-scale data ($W_k$) because \resp{the simulation results are reliable at least up to the particle Nyquist wavenumber (in Table \ref{table_nbody})} and the unaccounted baryonic effects can influence the small-scale results ($k \gtrsim 1 \, h  \, \Mpc^{-1}$). The fiducial cosmological model also received a high weighting ($W_c$)\footnote{
Accordingly, the weights were set to $W_z=8,3,1$ and $0.3$ for $z \leq 0.1, \, 0.1<  z \leq 1, \, 1 < z \leq 3$ and $z>3$, respectively; ~$W_k=3,1$ and $0.3$ for $k_{\rm max} / (h \, \Mpc^{-1}) \leq 3.2, \, 3.2< k_{\rm max} / (h \, \Mpc^{-1}) \leq10$, and $k_{\rm max}  / (h \, \Mpc^{-1}) >10$, respectively; and $W_c=1 \, (8 \times 10^{-4})$ in the \textit{Planck} 2015 model (otherwise). These values were chosen to achieve $10 \, \%$ accuracy of the fitting with $k < 3 \, h {\rm Mpc}^{-1}$ at $z=0\text{--}3$ in the \textit{Planck} 2015.}.

The analysis included all triangles $(k_1,k_2,k_3)$ satisfying the following three conditions: 
\begin{enumerate}
\item Relative standard deviation below $10 \, \%$ (i.e., $\Delta \! \ln \! B_{\rm sim} <0.1$).
\item Shot-noise contribution below $3 \, \%$.
\item If the deviation between the larger- and smaller-box simulation results exceeds $3 \, \%$ and the statistical error of $\Delta \! \ln \! B_{\rm sim}$ is below $3 \, \%$, we reject the larger-box result and use the smaller-box result only. In the larger-box simulation, the $B_{\rm sim}$ at high $k$ is reduced by the lack of spatial resolution (see also Figure \ref{fig_pk_compare} in the PS case).
\end{enumerate}
Conditions a) and b) exclude the data points at very small $k$ and large $k$, respectively. Condition c) negligibly affects the data selection.
Figure~\ref{fig_k1k2k3_region} plots the triangles $(k_1,k_2,k_3)$ satisfying the above three conditions for $L=1,2,4 \, h^{-1} \, \Gpc$ at $z=0.55$.
In the range $0.1 \lesssim k/(h \, \Mpc^{-1}) \lesssim 2$, the simulation results for all box sizes were overlapping and the fit was reliable.
As clarified in Figure~\ref{fig_k1k2k3_region}, the simulations covered almost {\it all} triangles up to $k=3 \, h \, \Mpc^{-1}$.
\resp{Note that most of the triangles were squeezed; the instances of equilateral and flattened cases were minor. Therefore, the fitting to squeezed cases is critically important.}
In the bottom panel, the discontinuity at $k_{\rm max} \simeq 3 \, (0.8) \, h \, \Mpc^{-1}$ for $L=1 \, (4) \, h^{-1}\, \Gpc$ can be explained by the box-size change from $L$ to $L/10$ $(L/4)$ when implementing the folding scheme (see also subsection~3.2).
The bottom panel is devoid of triangles in the lower right part, indicating that the calibration did not include very squeezed cases ($k_{\rm max} \gg k_{\rm min}$).
These cases lie outside the maximum $k_{\rm max}/k_{\rm min}$ ($512$), which is determined by the number of FFT grids ($1024^3$). 
The folding method does not change this ratio ($512$).
%This ratio ($512$) does not change after applying the folding method.
The number of independent triangular bins 
%\resp{(i.e., without overlapping among the different box sizes)}
calibrated in the simulations of each cosmological model was approximately $950$ at low $z$ ($z=0, 0.55, 1$ and $1.48$) and $690$ at high $z$ ($z=2,3$ and $5$), respectively.

\resp{Figure~\ref{fig_kmin_kmax} plots the maximum and minimum wavenumbers in the calibration. The minimum $k$ of simulation is larger at higher $z$ because the relative error $\Delta \! \ln \! B_{\rm sim}$ is larger. The maximum $k$ decreases at higher $z$ because the shot noise is not negligible at small scales.  Note again that all the triangles in this $k$ range are not included in the calibration (e.g., very squeezed cases are missing; see also Figure \ref{fig_k1k2k3_region}).}

%The total number of triangles (including overlapping among the different box sizes) are huge : $\sim 2.1 \times 10^4$ for the Planck2015 and $\sim 5.2 \times 10^5$ for the $40$ models.
%The average number in each redshift is $\sim 1600 \, (5.2 \times 10^4)$ for the Planck2015 (others in total) at $z=0-1.48$.

For a fair comparison, the simulation results and the fitting formula predictions should be binned consistently because the BS is sensitive to the binning, especially at the squeezed limit \citep{Sefusatti2010,Namikawa2019}.
Throughout this paper, the binned fitting was computed as
\begin{align}
 B^{\rm bin}(k_1,k_2,k_3) =& \frac{1}{N_{\rm triangle}} \int_{|\bfk^\prime_1| \in k_1} \hspace{-0.5cm} d^3 k_1^\prime \int_{|\bfk^\prime_2| \in k_2} \hspace{-0.5cm} d^3 k_2^\prime \int_{|\bfk^\prime_3| \in k_3} \hspace{-0.5cm} d^3 k_3^\prime 
 \nonumber \\
 &~~ \times B(k^\prime_1,k^\prime_2,k^\prime_3)  \, \delta_{\rm D} (\bfk^\prime_1+\bfk^\prime_2+\bfk^\prime_3),
\label{binned_BS_fit}
\end{align}
where $B(k_1^\prime,k_2^\prime,k_3^\prime)$ is the unbinned fitting and the number of triangles is
\begin{align}
N_{\rm triangle}=\int_{|\bfk^\prime_1| \in k_1} \hspace{-0.5cm} d^3 k_1^\prime \int_{|\bfk^\prime_2| \in k_2} \hspace{-0.5cm} d^3 k_2^\prime \int_{|\bfk^\prime_3| \in k_3} \hspace{-0.5cm} d^3 k_3^\prime \, \delta_{\rm D} (\bfk^\prime_1+\bfk^\prime_2+\bfk^\prime_3).
\end{align}
Here $k_{\rm i}$ is the weighted mean wavenumber, defined as $k_{\rm i} = \int_{|\bfk^\prime_{\rm i}| \in k_{\rm i}} \!\! d^3 k_{\rm i}^\prime \, k_{\rm i}^\prime \times [\int_{|\bfk^\prime_{\rm i}| \in k_{\rm i}} \!\! d^3 k_{\rm i}^\prime \,]^{-1}$.
The effect of the binning on BS is shown in Figure \ref{fig_bk_1-3haloterms_bin-unbin}. 
Note that although the unbinned triangle ($k^\prime_1,k^\prime_2,k^\prime_3$) satisfies the triangle condition (i.e., $|k^\prime_1-k^\prime_2|<k^\prime_3<k^\prime_1+k^\prime_2$), the bin center ($k_1,k_2,k_3$) may violate this condition.

\resp{We now comment on the effect of bin width $\Delta k$ on the calibration result. A finer bin width reduces the binning uncertainty and improves the calibration, at the cost of increased sample variance (as $\Delta B_{\rm sim} \propto \Delta k^{-3/2}$ under the Gaussian approximation). Therefore, the appropriate $\Delta k$ is not easily interpreted. Because binning smooths out the fine-$k$ BS features over the bin width, the accuracy of our fitting formula may be degraded if the user adopts a finer $\Delta k$ than ours. In subsection 4.3, we will check the bin width dependence by comparing $\Delta \log_{10} k=0.1$ and $0.05$ in the \textit{Planck} 2015, and confirm the agreement of the two results.}

\subsection{Fitting to perturbation theory}

As the simulation result is noisy at large scales, the calibration on the linear to quasi-linear scales was also performed by perturbation theory.
\resp{The same approach was adopted by \cite{SA2019} for modeling the non-linear matter PS.}
Here we applied one-loop standard perturbation theory (SPT) which includes the tree level and the next-to-leading-order terms \citep[e.g.,][]{Scocci1997,Scocci1998}.
The chi-square was defined analogously to Eq.~\eqref{chi_sim}:
\begin{align}
  \chi_{\rm spt}^2(\bfp) &= W_{\rm spt} \resp{ \sum_{c=1}^{41} \sum_{z=0}^{10} \sum_{k_1,k_2,k_3}^{0.3\,h\,\mathrm{Mpc}^{-1}}} W_z W_c \nonumber \\
  \times & \left[ \frac{B(k_1,k_2,k_3;\bfp) - B_{\rm spt}(k_1,k_2,k_3)}{\Delta B(k_1,k_2,k_3) + \epsilon(k_1,k_2,k_3)} \right]^2,
  \label{chi2_spt}
\end{align}
where $B_{\rm spt}$ is the SPT prediction. Note that $B$ and $B_{\rm spt}$ require no binning in this case.
We set $\Delta B = 0.5 \, |B_{\rm spt}-B_{\rm tree}|$, and $\epsilon=0.01 \times B_{\rm spt}$.
We also set $W_{\rm spt}=0.08$ to bias the simulation calibration\footnote{The resulting $\chi^2_{\rm sim}$ was approximately $40$ times larger than $\chi^2_{\rm spt}$.}, $W_c=1 \, (3 \times 10^{-5})$ in the \textit{Planck} 2015 (otherwise), and $W_z$ as prescribed in subsection~4.1.
All triangles ($k_1,k_2,k_3$) satisfying that $B_{\rm spt}$ agrees with $B_{\rm tree}$ within $5 \, \%$ and up to $0.3 \, h \, \Mpc^{-1}$ were included, thus restricting the fitting to large scales.
\resp{In the low-$k$ limit, the $\epsilon$ term dominates the denominator of Eq.~(\ref{chi2_spt}). As $k$ approaches $0.3 \, h \, \Mpc^{-1}$, the $\Delta B$ term ($<0.025 \times B_{\rm spt}$) dominates.}
%Although any unbinned triangular configurations are valid in the SPT calibration,
We used the central bin values of $(k_1,k_2,k_3)$ which were also used in the $L=4 \, h^{-1} \, \Gpc$ simulation with bin width $\Delta \log_{10} k =0.1$.
%Then, the calibration covers a range of $1.6 \times 10^{-3} \leq k/(h {\rm Mpc}^{-1}) < 0.3$ at $z=0\text{--}10$.
Figure~\ref{fig_k1k2k3_region} shows all triangles used in the \resp{SPT} calibration \resp{at $z=0.55$} (orange diamonds).
The average number of triangles was $350$ in each cosmological model at each redshift.
\resp{Figure~\ref{fig_kmin_kmax} shows the maximum and minimum $k$ in the SPT calibration. The minimum $k$ is $1.6 \times 10^{-3} \, h \, \Mpc^{-1}$ for all the redshifts. The maximum $k$ slightly increases from $0.18$ to $0.28 \, h \, \Mpc^{-1}$ from low to high $z$, because the SPT approaches the tree level at higher $z$.}
%(recall that $B_{\rm spt}$ and $B_{\rm tree}$ agree within $5 \, \%$ in the calibration).}

%The SPT prediction is more reliable even at higher $k$ for higher $z$.
%$\sim 4900 \, (1.9 \times 10^5)$ for the Planck2015 (others in total). 

\subsection{Results}

\begin{figure*}
\begin{center}
\includegraphics[width=15cm]{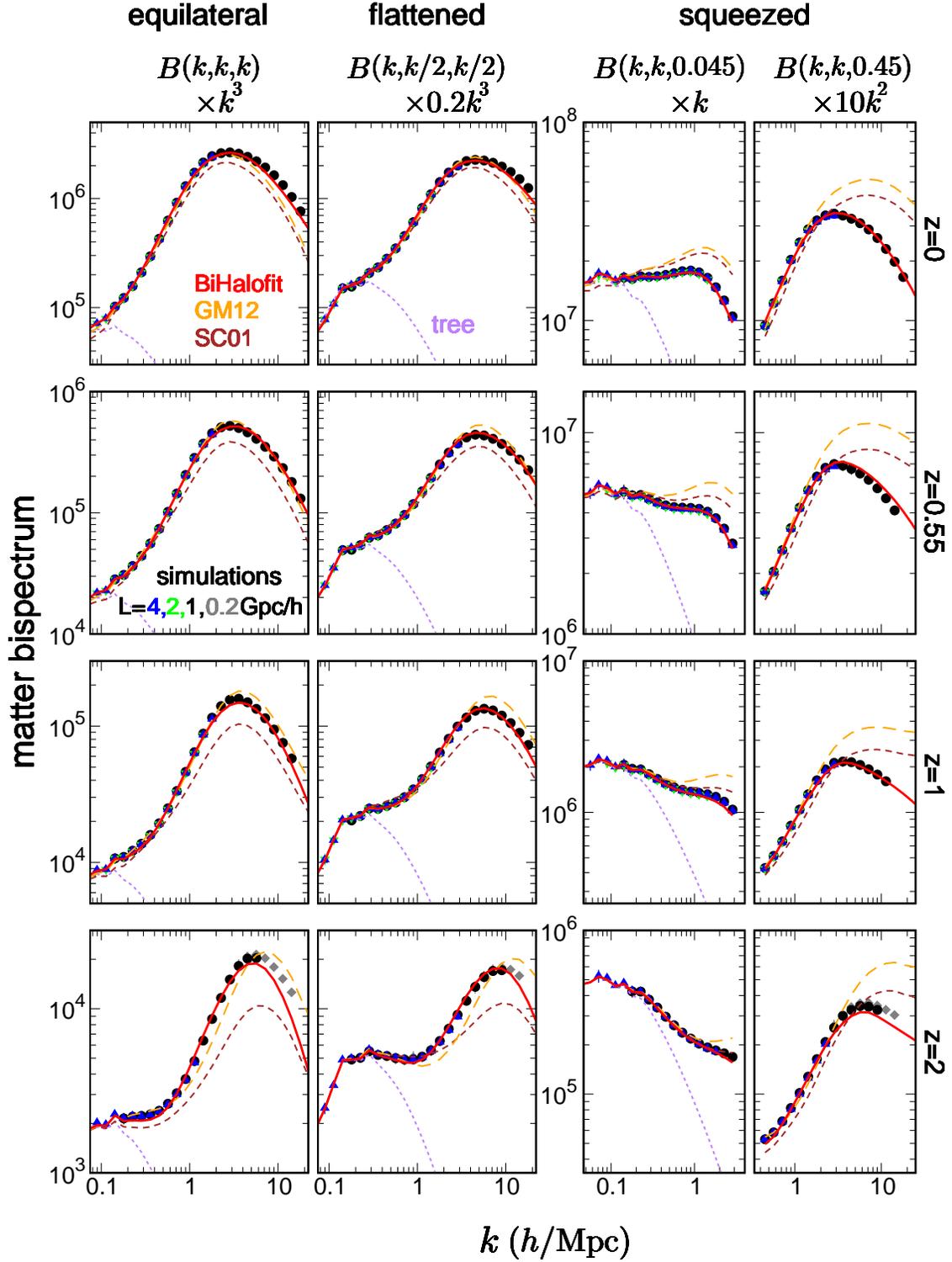}
\end{center}
\vspace*{-0.2cm}
\caption{Comparison of matter bispectra obtained in $N$-body simulations and the fitting formulas for the \textit{Planck} 2015 best-fit $\Lambda$CDM model. The curves denote the theoretical models: our fit (BiHalofit; solid red), \cite{GM12} (GM12; long-dashed orange), \cite{SC01} (SC01; short-dashed \respp{brown}) and the tree level (dotted purple).
Symbols denote the simulation results in various box sizes: $L=4,2,$ and $1 \, h^{-1} \, \Gpc$ (blue triangles, green triangles, and black circles) and $L=200 \, h^{-1} \, \Mpc$ (gray diamonds). 
The theoretical and simulation results are consistently binned (bin width  $\Delta \log_{10} k=0.1$).
Along the vertical axis, the bispectrum is multiplied by $\propto k^n$ ($n=1,2$ or $3$), as denoted above the top panels to clarify the presentation.
Throughout this paper, the units of $k$ and $B(k_1,k_2,k_3)$ are $h \, \Mpc^{-1}$ and $(h^{-1} \, \Mpc)^6$, respectively.
\label{fig_bk_4configs}
}
\end{figure*}

\begin{figure*}
\begin{center}
\includegraphics[width=15cm]{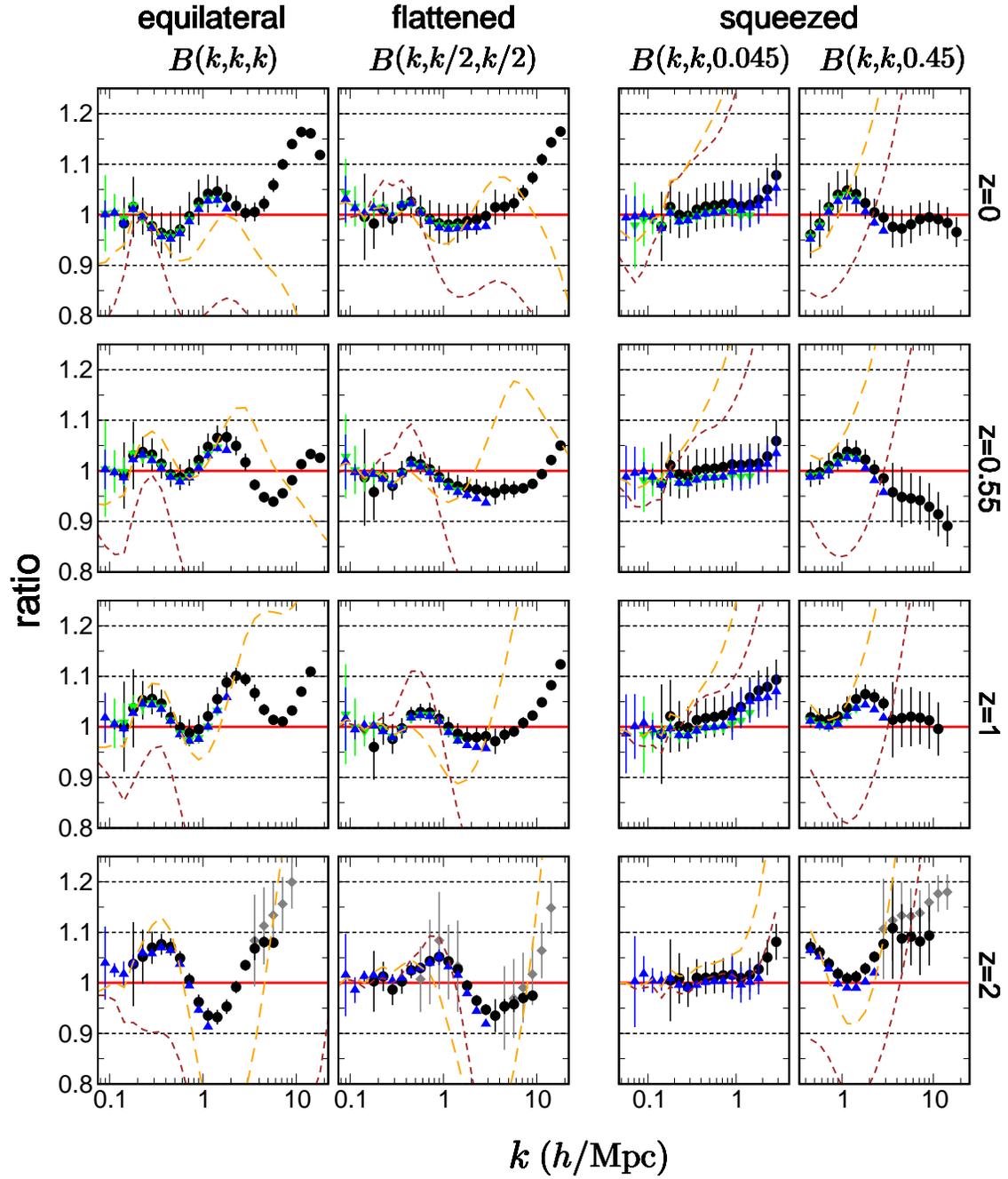}
\end{center}
\vspace*{-2cm}
\caption{Same as Fig.~\ref{fig_bk_4configs}, but relative to the red curves (BiHalofit). 
\label{fig_bk_4configs_residual}
}
\end{figure*}

\begin{figure*}
\begin{center}
\includegraphics[width=15cm]{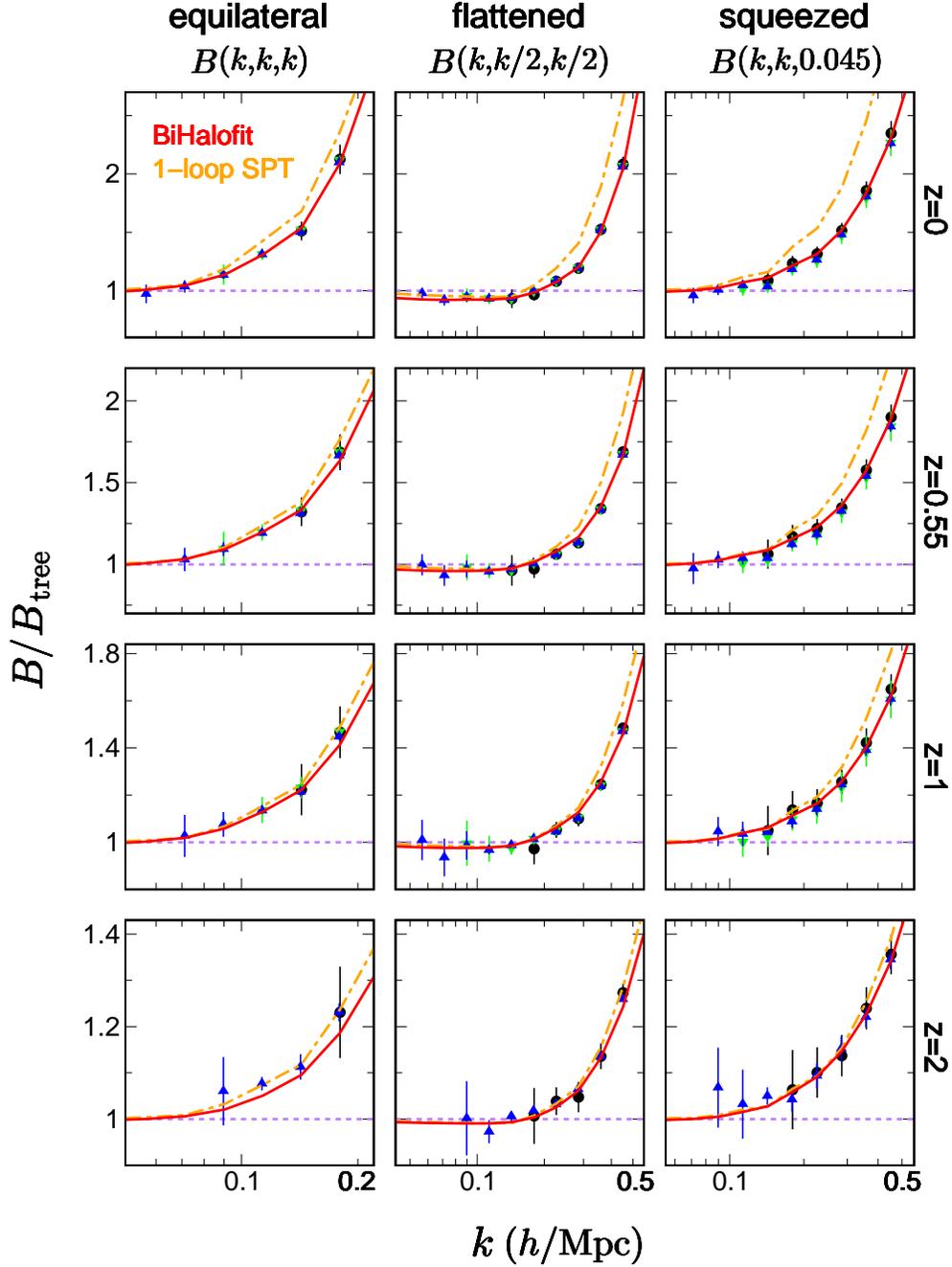}
\end{center}
\vspace{-2cm}
\caption{Similar to Fig.~\ref{fig_bk_4configs_residual}, but relative to the tree-level results on quasi-linear scales. The dot-dashed orange curves are the binned one-loop SPT predictions.}
\label{fig_bk_ratio-tree}
\end{figure*}

\begin{figure*}
\begin{center}
\includegraphics[width=15cm]{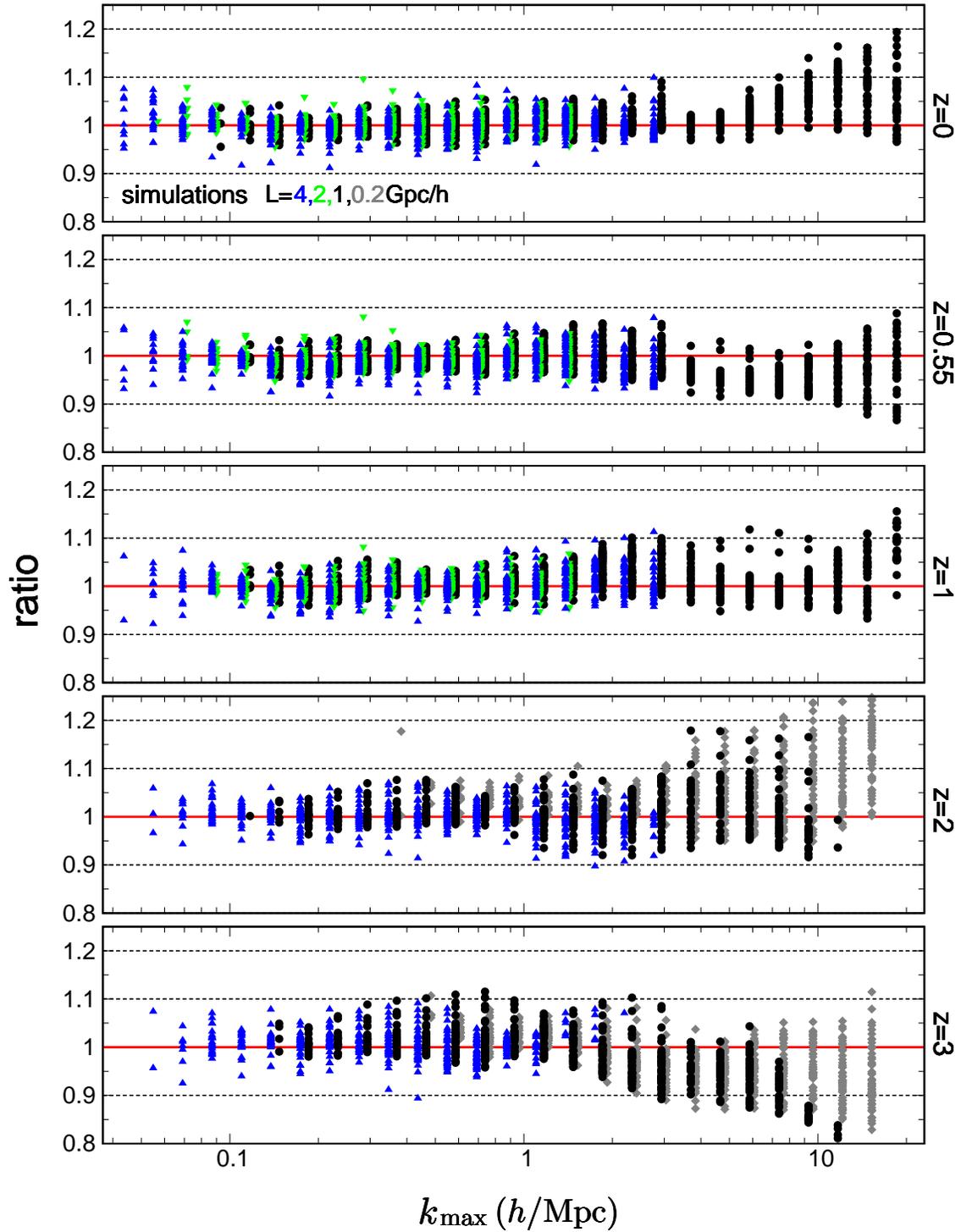}
\end{center}
\vspace{-1cm}
\caption{Bispectrum ratios of simulation results to BiHalofit for {\it all} triangles in the \textit{Planck} 2015 model. The horizontal axis presents the maximum wavenumber of ($k_1,k_2,k_3$).
Dozens of points are distributed along the vertical axis at each $k_{\rm max}$.
The horizontal positions for different $L$ are slightly offset to clarify the presentation. 
\label{fig_bk_allconfigs_residual}
}
\end{figure*}

\begin{figure*}
\begin{center}
\includegraphics[width=15cm]{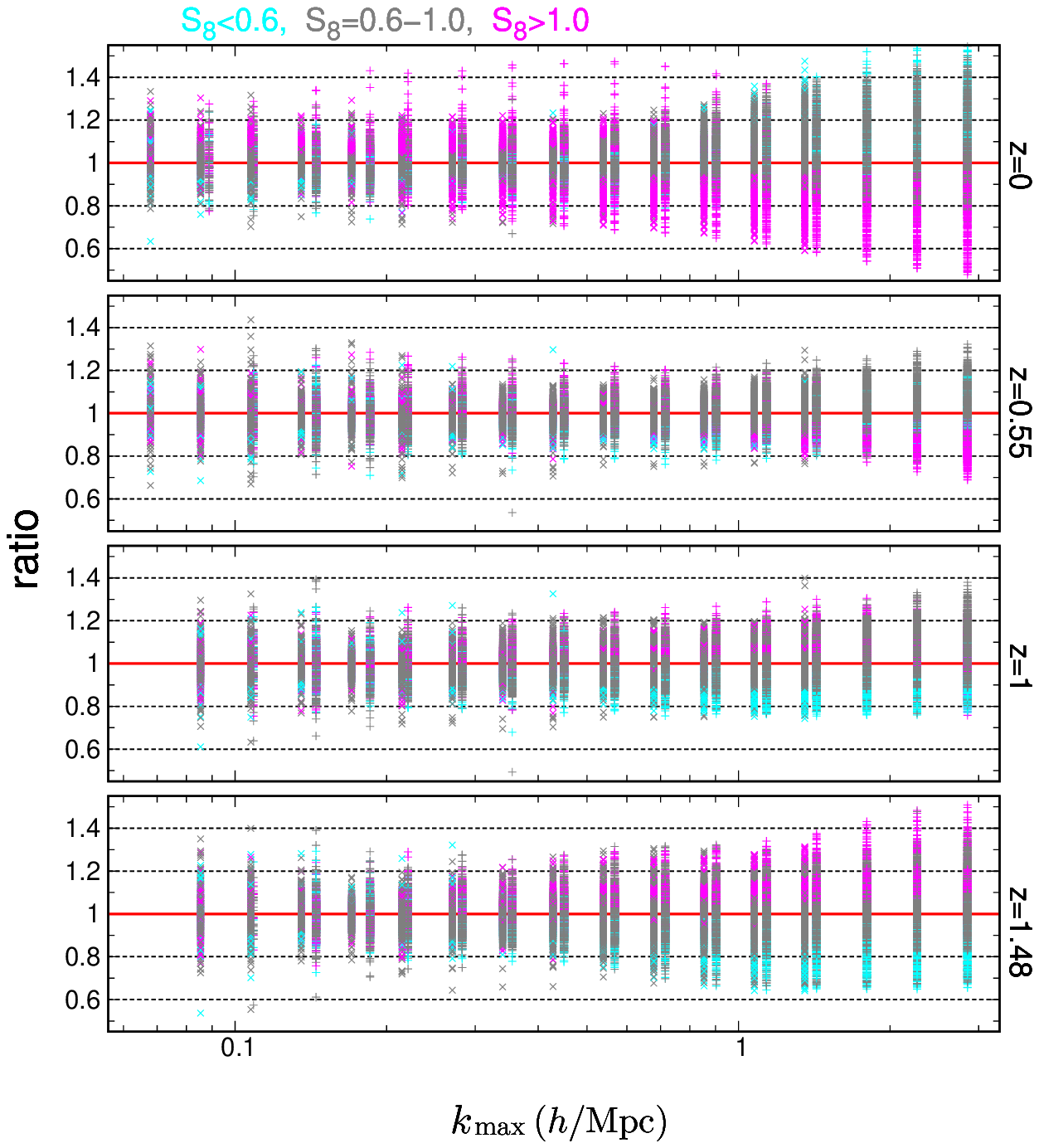}
\end{center}
\vspace{-4.3cm}
\caption{Same as Fig.~\ref{fig_bk_allconfigs_residual}, but showing the results of the $40$ $w$CDM models with $z=0\text{--}1.48$. The colors correspond to the cosmological models shown in Fig.~\ref{fig_cosmo_params}. The cyan, gray and magenta points were obtained in models with different ranges of $S_8 \equiv \sigma_8 (\Omega_{\rm m}/0.3)^{0.5}$.  The plus signs (crosses) are the simulation results of $L=1 \, (2) \, h^{-1} \, \Gpc$. Note that every data point has an intrinsic scatter of $\sim 10 \, \%$ because there is a single realization in each $w$CDM model.}
\label{fig_bk_allconfigs_residual_40models}
\end{figure*}

The total $\chi^2$ was computed as
\beq
 \chi^2(\bfp) = \chi_{\rm sim}^2(\bfp) + \chi_{\rm spt}^2(\bfp).
\eeq
The best-fitting parameters $\bfp$ were then numerically searched by minimizing $\chi^2$.
The resulting best-fit model is presented in Appendix~B.
The minimum was found by the downhill simplex routine (amoeba) in Numerical Recipes \citep{NumericalRecipes}.

Figure~\ref{fig_bk_4configs} plots the matter BSs computed by the tree-level formula, our fitting formula, SC01, and GM12, along with the simulation results of the \textit{Planck} 2015 model with $z=0\text{--}2$.
From left to right, the four panels correspond to particular triangle configurations: equilateral (i.e., $k_1=k_2=k_3$), flattened ($k_1=2k_2=2k_3$) and two squeezed cases ($k_1=k_2 \gg k_3$ with $k_3=0.045$ and $0.45 \, h \, \Mpc^{-1}$). 
In this and the following figures, the simulation data points satisfy conditions a) -- c) in subsection~4.1.
In the fitting formulas of SC01 and GM12, we applied the measured PS \resp{of the simulations} to remove the inaccuracy of the PS appearing in these models.
Figure~\ref{fig_bk_4configs_residual} plots the ratios of SC01, GM12 and the simulation results to our fitting formula results.
Clearly, our fitting formula agreed with the simulations over the tested scales, redshifts, and triangle shapes.
In contrast, the previous formulas over-predicted the squeezed BS, as previously reported by \citet{Namikawa2019}.
%Therefore, the deviation is caused by
The simulations performed in different box sizes were also consistent.

Figure~\ref{fig_bk_ratio-tree} shows the ratios of the modeled and simulated BSs to the tree-level BS on quasi-nonlinear scales.
On larger scales, both our simulations and fitting formula were consistent with the tree-level prediction.
Meanwhile, the one-loop SPT slightly over-predicted the BS on quasi-nonlinear scales at low $z$ $(z<1)$ \cite[consistent with Figure 19 of][]{Lazanu2016}, but its inaccuracy improved at higher redshifts.
In the flattened case, the SPT slightly suppressed the BS at $k \sim 0.1 \, h \, \Mpc^{-1}$ and our model captures this trend.
Some data points at $k \, <0.1 \, h \, \Mpc^{-1}$ were omitted because their relative error exceeded $10 \, \%$.
\resp{The rms deviation between the formula and SPT was $0.96 \%$ for all the triangles in our sample of $z=0\text{--}10$. Therefore, the accuracy reached the percent level at largest scales ($k<0.3 \, h \, \Mpc^{-1}$).}

Figure~\ref{fig_bk_allconfigs_residual} shows the BS ratios of the simulation results to our formula for {\it all} triangles satisfying conditions a) -- c) in subsection 4.1.
There are approximately $1800$ data points in each redshift.
Our model agreed with the simulations within $10 \, (15) \%$ up to $k=3 \, (10) \, h \, \Mpc^{-1}$ for $z=0\text{--}3$.
At $z=5 \, (10)$, the agreement was $20 \, \%$ up to $k=3 \, (1) \, h \, \Mpc^{-1}$.
The rms deviation was $2.7 \, (3.2) \%$ \resp{and $3.7 \, (5.0) \%$} up to $k=3 \, (10) \, h \, \Mpc^{-1}$ for $z=0\text{--}3$ \resp{and $z=0\text{--}10$, respectively}.
Moreover, the accuracy was independent of bin width, as confirmed by setting a narrower bin width ($\Delta \log_{10} k=0.05$) in the same tests.
The narrow bins yielded an rms deviation of $2.9 \, (3.4) \%$ up to $k=3 \, (10) \, h \, \Mpc^{-1}$ at $z=0\text{--}3$, quantitatively consistent with the above results.
\resp{Therefore, the accuracy is approximately $3 \, \%$ at $k<10 \, h \, \Mpc^{-1} $ and $z=0\text{--}3$ for most of the triangles (but it reaches $10\text{--}15 \%$ in the worst cases).}

%Mention the case for smaller bin-width $\Delta \log k =0.05$.

Figure~\ref{fig_bk_allconfigs_residual_40models} plots the BS ratios of the simulations to our formula in the $40$ $w$CDM models.
In this case, as we prepared a single realization for each cosmological model, the BS measurements had a relatively large scatter (typically $10 \%$). 
All data points satisfied conditions a) -- b) in subsection~4.1.
There are a huge number of data points $(\sim 5 \times 10^4)$ at each redshift.
The rms deviation was $8.0 \, (11.2) \%$ up to $k=3 \, (10) \, h \, \Mpc^{-1}$ for $z=0\text{--}1.5$. 
\resp{The deviation includes the $10 \%$-level sample variance of the simulations.}
%The cosmological models are divided to three groups in terms of $S_8$.

To further investigate the cosmological dependence of the accuracy, we divided the models into three groups with different ranges of $S_8=\sigma_8(\Omega_m/0.3)^{0.5}$. The data points shown in Figure \ref{fig_bk_allconfigs_residual_40models} are color-coded as described in the caption of Figure \ref{fig_cosmo_params}.
Our formula agreed with the simulations within $\sim 20 \, \%$ for $S_8=0.6\text{--}1.0$, but the agreement degraded outside this $S_8$ range because the fluctuation amplitude ($\sigma_8$) and the linear growth factor largely differed between these models and the \textit{Planck} 2015 model.
As all cosmological models converged to the Einstein--de Sitter model at high $z$, the fits improved at higher redshifts.

\resp{In the $40$ $w$CDM models, the rms deviation between the formula and the SPT is $1.3 \%$ at $k<0.3 \, h \, \Mpc^{-1}$ and $z=0\text{--}10$. Therefore, our formula is well consistent with the SPT at largest scales.}

\section{Baryonic effects}

\begin{figure*}
\begin{center}
\includegraphics[width=18cm]{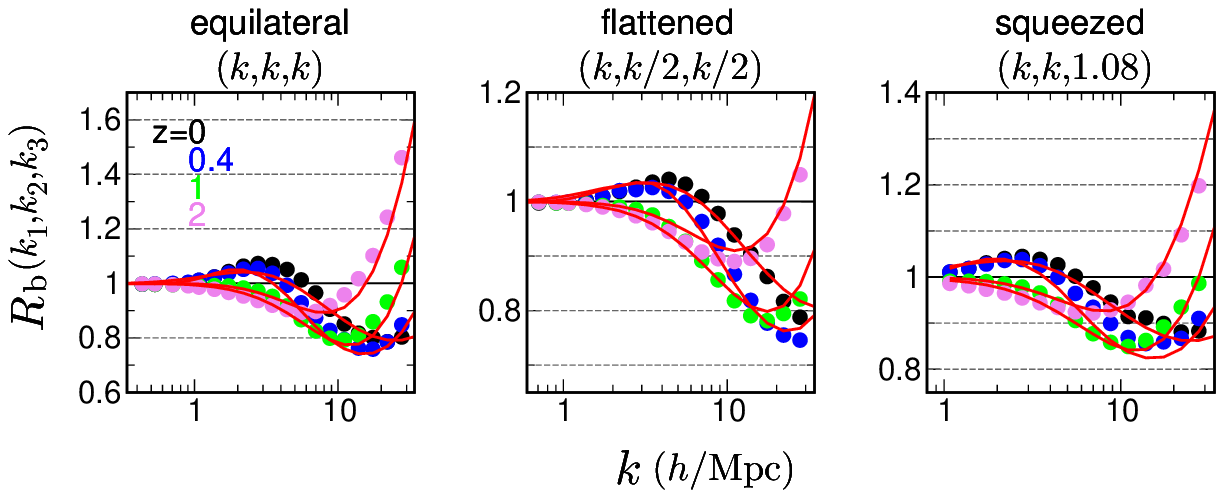}
\end{center}
\caption{Ratio of bispectrum with to without baryons, $R_{\rm b}=B_{\rm b}/B_{\rm dmo}$ defined in Eq.~\eqref{ratio_baryon}, measured from the TNG300-1. The filled circles are the total-matter (dark matter and baryons) bispectrum divided by that from the dark-matter-only run. The red curves are our fit given in Appendix~C. \respp{We comment that $R_{\rm b}$ varies by $\sim 10 \, \%$ among hydrodynamical simulations performed in different groups, because the baryonic feedback models differ (see the main text).}
}
\vspace{0.3cm}
\label{fig_bk_baryon_ratio}
\end{figure*}

\begin{figure}
\begin{center}
\includegraphics[width=8cm]{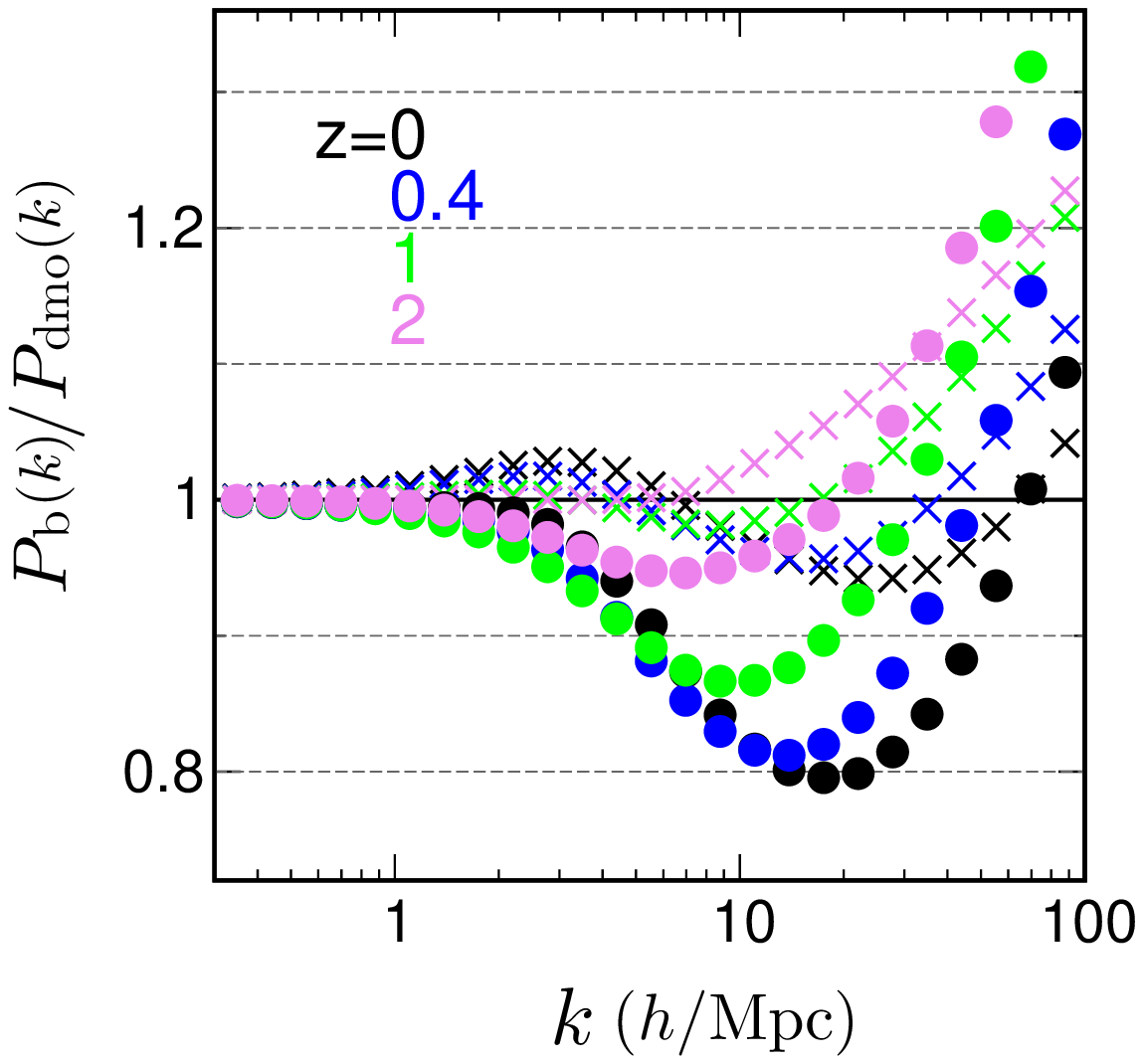}
\end{center}
\vspace*{-0.3cm}
\caption{Similar to Fig.~\ref{fig_bk_baryon_ratio}, but ratio of $P(k)$ measured from the TNG300-1.  
The filled circles (crosses) are the total-matter (dark-matter) $P(k)$ \resp{from the hydrodynamic run} divided by that from the dark-matter only run. 
}
\label{fig_pk_baryon_ratio}
\end{figure}

Our $N$-body simulations did not include the baryonic processes such as gas cooling, star formation, supernovae and active galactic nucleus (AGN) feedbacks. Baryons are known to significantly affect the nonlinear PS at $k \gtrsim 1 \, h \, \Mpc^{-1}$ \citep[e.g.,][]{vanD2011,Sembo2011b,Osato2015,Hellwing2016,Chisari2018,Chisari2019}.
In this section, the baryonic effects on the BS fitting formula are investigated in state-of-the-art hydrodynamic simulations using the IllustrisTNG data set\footnote{\url{http://www.tng-project.org}} \citep{Marinacci2018,Naiman2018,Nelson2018a,Nelson2018,Pillep2018,Springel2018}.
The simulations incorporate astrophysical processes in a subgrid model and thereby follow the galaxy formation and evolution processes. 
The IllustrisTNG project conducted three sets of simulations in different box sizes, with three mass resolutions in each box size.
Here we used the highest-resolution simulation in the largest box (referred to as TNG300-1) of size $L=205 \, h^{-1} \, \Mpc \, (\simeq 300 \, \Mpc)$. 
This box contains $2500^3$ dark matter particles and the same number of baryon particles.
The cosmological model of IllustrisTNG is based on the \textit{Planck} 2015 best-fit $\Lambda$CDM \citep{Planck2015}.
The collaboration has released the particle positions and masses of dark matter and baryons (in the forms of gas, stars and black holes) at $z=0\text{--}20$.
The IllustrisTNG team also performed dark-matter-only (dmo) runs.
By comparing the simulations in the presence and absence of baryons, we can single out the impact of baryons on matter clustering. 

To calculate the density contrast, we assigned the particle masses to $1024^3$ grid cells and measured the BS as described in subsection~3.4.
\resp{The bin width was set to $\Delta \log_{10}k=0.1$.}
We calculated the BS ratio of the simulations with baryons ($B_{\rm b}$) to the dmo run ($B_{\rm dmo}$),
\beq
 R_{\rm b}(k_1,k_2,k_3) = \frac{B_{\rm b}(k_1,k_2,k_3)}{B_{\rm dmo}(k_1,k_2,k_3)}.
\label{ratio_baryon}
\eeq
We measured this ratio at \resp{eleven} redshifts: $z=0,0.2,0.4,0.7,1,1.5,2,3,5,\resp{7}$ and $\resp{10}$.

Figure~\ref{fig_bk_baryon_ratio} plots the ratios in Eq.~(\ref{ratio_baryon}) for three triangle configurations in the range of $z=0\text{--}2$.
To reduce the sample-variance scatter in the ratio at large scales, the simulations with and without baryons had the same seed in their initial conditions.
The baryons suppressed the BS amplitude at $k \sim 10 \, h \, \Mpc^{-1}$ by the AGN feedback but strongly enhanced the BS amplitude at high $k \, (> 10 \, h \, \Mpc^{-1})$ by the gas cooling.
This trend is consistent with the PS (see also Figure \ref{fig_pk_baryon_ratio}).
However, at intermediate scales ($k \simeq 1\text{--}10 \, h \, \Mpc^{-1}$) and low redshifts $(z \, <1)$, the baryons slightly {\it enhanced} the amplitude by $\sim 10 \, \%$.
To our knowledge, this small enhancement has not been commonly observed in PS.

Figure~\ref{fig_pk_baryon_ratio} plots the PS ratios with and without baryons, computed in TNG300-1.
The circles (crosses) are \resp{the PSs of the total matter (dark matter component only) in the hydrodynamic run divided by that in} the dmo run. 
At intermediate scales, the plots of the crosses were slightly enhanced, whereas those of the circles were not.
The same feature is mentioned in section~3 of \cite{Springel2018}, consolidating that the enhancement source is the dark matter component. Moreover, the dark matter PS and the total-matter BS are enhanced at almost the same wavenumbers.

During the preparation of this paper, \cite{Foreman2019} posted an arXiv paper concerning the baryonic effects on BS measured in hydrodynamic simulations (including TNG300-1).
They reported the same trend and clarified its cause.
At late times ($z < 1$), the AGN feedback becomes less effective and the expelled gas re-accretes into a halo.
Gas contraction then affects the dark matter distribution in the halo.
As the BS is more sensitive to dark matter than the PS (see their subsection~3.1.1), the enhancement \resp{at intermediate scales} appears only in BS.
By studying the baryonic effects on both PS and BS, one can discriminate among baryonic models \citep{Sembo2013,Foreman2019}.

To incorporate the baryonic effect in our BS model, we constructed a fitting function of the ratio $R_{\rm b}$ in Eq.~\eqref{ratio_baryon}.
The results are plotted as the solid red curves in Figure \ref{fig_bk_baryon_ratio}, and the functional form is given in Appendix~C.
This fitted the measurements within $7.3 \, (5.3) \, \%$ for $k<10 \, h \, \Mpc^{-1}$ at low (high) redshift, $z=0\text{--} 1 \, (1.5\text{--}\resp{10})$.
The rms deviation was $1.8 \, \resp{(2.9)} \, \%$ for $k<30 \, h \, \Mpc^{-1}$ at $z=0 \text{--} 1 \, \resp{(1.5\text{--}10)}$.
In this data fitting, approximately $760 \, (\resp{8300})$ triangles existed at each redshift (over the full range $z=0$\text{--}$\resp{10}$).
To include the baryonic effects, the user can simply multiply $R_{\rm b}$ by the BS fitting formula. 
The same approach was adopted by \cite{Harnois2015}, who studied the baryonic effects on PS.

\resp{We comment that the BS ratio $R_{\rm b}$ varies by approximately $10 \%$ among hydrodynamical simulations, because the baryonic feedback models differ. \cite{Foreman2019} measured the BS ratio $R_{\rm b}$ for the equilateral case in four simulations: IllustrisTNG, Illustris \citep{Vogel2014}, BAHAMAS \citep{McCarthy2017} and EAGLE \citep{Schaye2015}. They reported a $10$\text{--}$20 \%$ variation in the results for $k > 1 \,h \, \Mpc^{-1}$ and $z=0$\text{--}$3$. In Illustris and BAHAMAS, the small enhancement at intermediate scales ($k=0.1$\text{--}$1 \, h \, \Mpc^{-1}$; see Figure \ref{fig_bk_baryon_ratio}) was absent, but the suppression at small scales ($k \gtrsim 10 \, h \, \Mpc^{-1}$) was amplified because these models implemented a stronger AGN feedback than IllustrisTNG. Therefore, the uncertainty in our fitting formula also hovered around $10 \%$.}
%was calibrated from one model of them, we need further study of this topic.}

\section{Comparison with weak-lensing simulations} 

Using the fitting formula of matter BS calibrated over wide ranges of wavenumbers and redshifts, we can predict the lensing observables by integrating along the line of sight.
This section compares our theoretical prediction with the weak-lensing BS measured in ray-tracing simulations.
We consider the convergence BS in two cases: CMB lensing (subsection~6.1) and cosmic shear (subsection~6.2).

The convergence field is a dimensionless matter density integrated along the line of sight toward the source.
%and can be constructed via the lensed temperature/polarization map (CMB lensing) and the shear field (cosmic shear).
The convergence at angular position $\bftheta$ for a source distance $r_{\rm s}$ is given by \cite[e.g.,][]{BS2001}
\beq
 \kappa(\bftheta) = \int_0^{r_{\rm s}} \!\! dr \, W(r,r_{\rm s}) \, \delta(r \bftheta,r;z),
\eeq
with the weight function 
\beq
  W(r,r_{\rm s}) = \frac{3 H_0^2 \Omega_{\rm m}}{2 c^2} \frac{r \left( r_{\rm s} -r \right)}{a(r) \, r_{\rm s}},
\eeq
where $r \, (r_{\rm s})$ is the comoving distance (to the source) and $a(r)$ is the scale factor.
The convergence BS is
\beq
  B_\kappa(\ell_1,\ell_2,\ell_3) = \int_0^{r_{\rm s}} \!\! dr \frac{W^3(r,r_{\rm s})}{r^4} B\left( \frac{\ell_1}{r},\frac{\ell_2}{r},\frac{\ell_3}{r} ; z \right),
\label{bk_conv}
\eeq
where $\ell_i \, (= k_i r)$ is the multipole moment and $B(k_1,k_2,k_3;z)$ is the matter BS at $z$.
This formula was derived under the flat-sky and the Born approximations.
When the source has a high redshift, the Born approximation is less accurate and must be adjusted by post-Born corrections \citep{PL2016}.
These corrections are necessary only in CMB lensing \citep[in cosmic shear, their contribution is $\mathcal{O}(1 \, \%)$, see Figure 7 of][]{PL2016}.
%The post-Born BS depends on the matter PS \cite[see e.g.,][Eq. (11)]{Namikawa2019}. 

For a source with a given redshift, the convergence BS is more sensitive to lower-$z$ structures than the convergence PS  \citep[see, e.g., see Figure 4 of][]{TJ2002}, because the matter BS (PS) evolves proportionally to the fourth (second) power of the linear growth factor in the linear regime.
Therefore, the matter BS and PS can probe structures with different redshifts in a complimentary manner.

\subsection{CMB lensing}

\begin{figure*}
\begin{center}
\includegraphics[width=16cm]{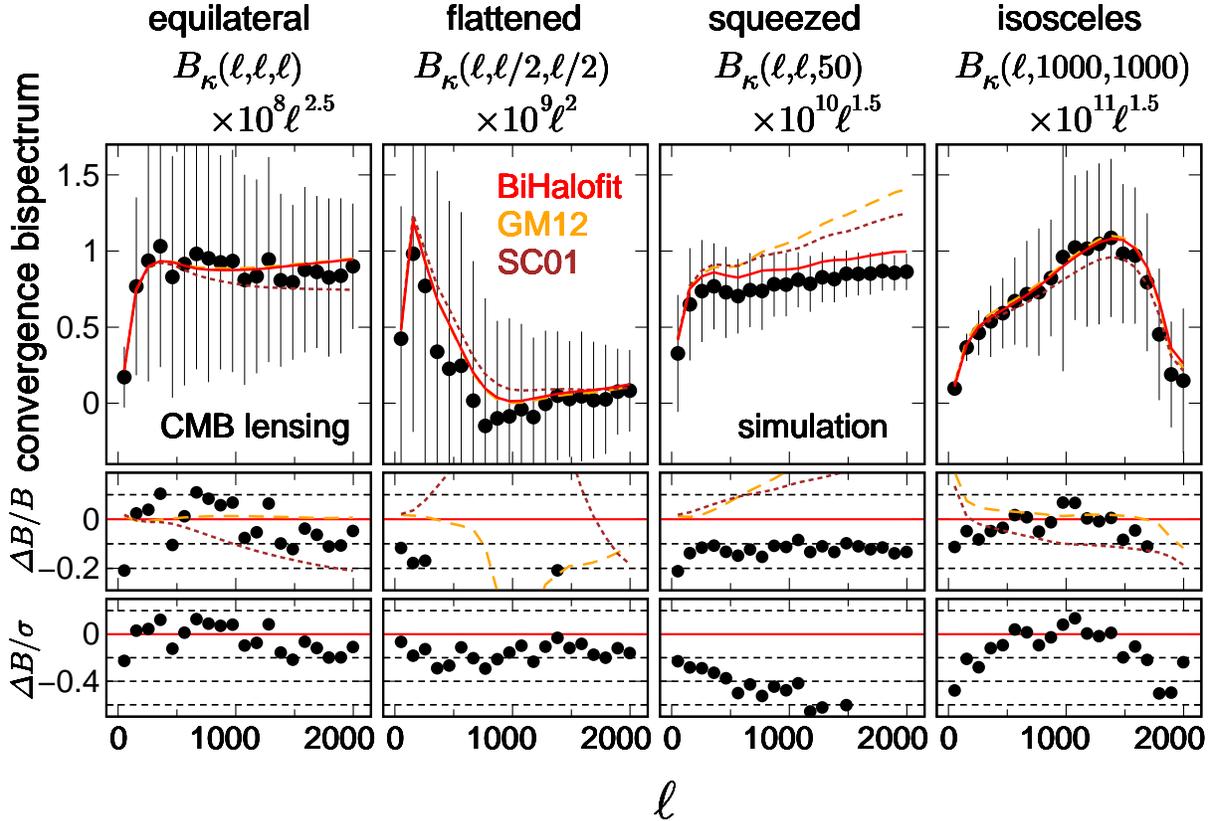}
\end{center}
\vspace{-0.cm}
\caption{Convergence bispectra measured from simulation maps of CMB lensing. The black symbols were averaged from $108$ full-sky maps \citep{Takahashi2017,Namikawa2019}. The error bars are the standard deviations scaled by $[({\rm survey~area})/(4 \pi)]^{-1/2}$. The solid red, dashed orange, and dotted \respp{brown} curves are the theoretical predictions based on BiHalofit, GM12, and SC01, respectively.
The middle panels plot the relative deviations from the red curves: $\Delta B/B \equiv B_\kappa/B_{\kappa}^{\rm BiHalofit}-1$. 
In the bottom panels, these deviations are further divided by the relative standard deviation ($\sigma/B$).
}
\vspace{0.3cm}
\label{fig_bk_cmb_lensing}
\end{figure*}

\cite{Namikawa2019} recently measured the convergence BS in full-sky light-cone simulations \citep{Takahashi2017}.
Here we compare their measurements with the theoretical predictions.
\cite{Takahashi2017} ran cosmological $N$-body simulations of the inhomogeneous mass distribution in the universe, from the present to the last scattering surface.
Their cosmological model was consistent with the WMAP 9yr result \citep{wmap9yr2013}.
The authors also calculated the light-ray paths deflected by the intervening matter in a ray-tracing simulation, which tracks the trajectories of the light rays emitted from the observer's position (at $z=0$) to the last scattering surface \citep[the ray-tracing scheme is detailed in][]{Shirasaki2015}.
Their results included the post-Born effects. 
\cite{Takahashi2017} provided $108$ full-sky convergence maps\footnote{These maps are available at \url{http://cosmo.phys.hirosaki-u.ac.jp/takahasi/allsky_raytracing}.} based on the \texttt{HEALPix} pixelization with $N_{\rm side}=8192 \, (4096)$, corresponding to a pixel size of $0.48 \, (0.96)$ arcmin \citep{Gorski2005}.
They confirmed that the convergence PS agrees with the theoretical CAMB prediction using the Halofit PS option (within $5 \, \%$ at $\ell \leq 2000$ on the high-resolution maps with $N_{\rm side}=8192$).

Figure~\ref{fig_bk_cmb_lensing} plots the BS measurements obtained from the $108$ maps with $N_{\rm side}=8192$ \citep{Namikawa2019}.
The theoretical predictions were computed for the WMAP 9yr cosmological model to be consistent with the simulations.
\resp{Here the nonlinear PS for GM12, SC01 and the post-Born correction was computed by the revised Halofit.}
For a fair comparison, both the theoretical predictions and simulation results were binned with the same bin width ($\Delta \ell=100$). 
The error bars were computed for the ideal full-sky measurement (i.e., the cosmic-variance limit) and scaled as $[({\rm survey~area})/(4 \pi)]^{-1/2} \Delta \ell^{-3/2}$, assuming Gaussian variance.
%In this and the next figures, the error bar denotes the standard error of the mean (i.e., the standard deviation divided by $108^{1/2}$). 
Overall, our fitting formula better predicted the BS of CMB lensing than the previously proposed formulas. In the equilateral case, the analytical and simulated BS agreed within $\sim 10\%$ on most angular scales. 
The differences were within $0.2 \sigma$ (bottom panels of Figure \ref{fig_bk_cmb_lensing}).
In the flattened case, the ratio (middle panel) was far from unity because the BS approaches zero at $\ell \gtrsim 1000$. 
This discrepancy is approximately $0.2\sigma$ of the cosmic variance.
In the squeezed and isosceles configurations, our fitting formula significantly reduced the discrepancy between the simulation result and the analytical prediction. 
%As clearly seen in the figure, our fitting formula agrees with the simulation well within about $10 \%$ accuracy.
%For the squeezed case, our formula gives much better agreement compared to previous fitting models but still shows a slight overestimation.

%The discrepancy between the analytic and simulated BS is significantly reduced by using our fitting formula compared to the use of the existing formulas, but is still non-zero. 
Although our fitting surely improved the prediction accuracy, noticeable discrepancies from the simulations were introduced by several sources.
First, the finite thickness of the lens planes employed in the ray-tracing simulations may affect the simulations at $\ell < 200$ \citep[the same effect on convergence PS is demonstrated in Figure 10 of][]{Takahashi2017}. 
Second, the flat-sky formula in Eq.~(\ref{bk_conv}) is inaccurate at large angular scales \citep[the accuracy of the flat-sky approximation in the cosmic-shear PS is detailed in][]{Kilbin2017,Kitching2017}. 
For example, in the squeezed limit, the minimum multipole is fixed as $\ell_3=50$, but a larger $\ell_3$ can mitigate the discrepancy \citep{Namikawa2019}. 
Since reducing the finite thickness of lens planes requires more numerically expensive simulations, we will leave the detailed study for future work. 

\subsection{Cosmic shear}

\begin{figure*}
\begin{center}
\includegraphics[width=16cm]{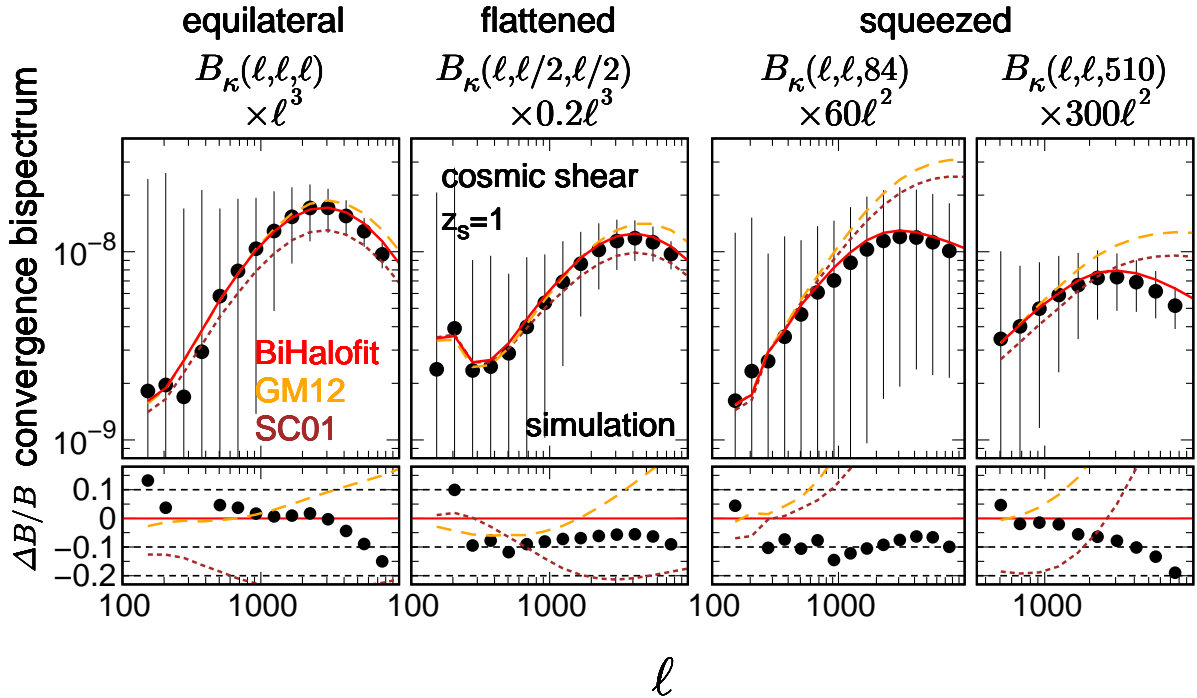}
\end{center}
\vspace{-1.5cm}
\caption{Convergence bispectra measured from $1000$ simulation maps at the source redshift $z_{\rm s}=1$ \citep{Sato2009,Kayo2013}. Black circles are the measured averages, and the red curves were predicted by our model. 
\resp{The dashed orange and \respp{brown} dotted curves were obtained by GM12 and SC01, respectively.}
The field of view of each map is $5 \times 5 \, {\rm deg}^2$ and the error bars are the standard deviations scaled by $[({\rm survey \, are})/(25 \, {\rm deg}^2))]^{-1/2}$.
The bottom panels plot the relative deviations from the red curves.  
}
\vspace{0.3cm}
\label{fig_bk_cosmic_shear}
\end{figure*}

Let us now consider the cosmic-shear signals in galaxy-shape measurements, which probe lower redshifts than CMB lensing.
\cite{Sato2009} ran cosmological $N$-body simulations and subsequent ray-tracing simulations under the flat-sky approximation.
Despite their small field of view ($5 \times 5 \, {\rm deg}^2$), they acquired sufficiently many weak-lensing maps ($1000$) for an accurate BS measurement.
Their cosmological model was consistent with the WMAP 3yr result \citep{Spergel2007}.

Figure~\ref{fig_bk_cosmic_shear} plots the convergence BS at a source redshift of $z_{\rm s}=1$, measured from the $1000$ maps by \cite{Kayo2013}.
The theoretical and simulation results were consistently binned with $\Delta \log_{10} \ell = 0.13$.
The simulation results were valid (within $5 \%$ error) up to $\ell \simeq 4000$, as confirmed by comparing the results with those from low- and high-resolution maps \citep{Sato2009,Valag2012}.
\resp{Overall, the plot shows a similar trend to the matter BS at $z=0.55$ (see Figure \ref{fig_bk_4configs}).} 
Our fitting formula well agreed with the simulation (within $10 \, \%$ level up to $\ell = 4000$).
The deviation at small scales $(\ell \gtrsim 4000)$ was attributed to the limited resolution of the simulation.

\resp{As the cosmological models in this and the previous subsections (6.1 and 6.2) differ from the \textit{Planck} 2015, the agreement %s
with the weak-lensing simulations provides a nontrivial validation of our formula for other cosmological models.}

\section{Discussion}

\subsection{Systematics in CMB lensing}

In CMB lensing measurements, the lensing map is reconstructed through mode-mixing of the CMB anisotropies induced by lensing \citep{HuOkamoto:2001}. Therefore, any other sources of mode-mixing can bias the lensing measurements and hence the BS of CMB lensing.
Bias can be sourced from instrumentation factors such as masking, inhomogeneous noise, beam, and point sources \citep{Hanson:2009:noise,Namikawa:2012:bias-hardening}, and from extragalactic foregrounds such as the thermal Sunyaev-Zel'dovich effect, the cosmic infrared background \citep{Osborne:2013nna,vanEngelen:2013rla,Madhavacheril:2018bxi}, and its lensing \citep{Mishra:2019qyd}. 
Calibration uncertainties in the CMB map are also important sources of systematic error, because when the lensing reconstruction is performed by a quadratic estimator, the measured BS depends on the sixth power of the map-calibration uncertainties. 
In contrast, the lensing PS depends on the fourth power of the map-calibration uncertainties.
Combining the BS and PS is expected to constrain the bias contributed by the instrumental uncertainties and astrophysical sources, because these sources affect the spectra in different ways. A joint analysis of the PS and BS is therefore crucial for a robust cosmological analysis in future CMB experiments.

\subsection{Intrinsic alignment}

The intrinsic alignment (IA) of galaxies is a major systematic error in cosmic shear \citep[reviewed by][]{Troxel2015,Joachimi2015}.
A massive structure near the source galaxy exerts a tidal force that distorts the shapes and contaminates the lensing signal.
Approximately $10 \, \%$ of the cosmic-shear BS is contaminated by this mechanism \citep{Sembo2008}. 
Several authors have proposed methods for mitigating or removing the contamination from the signal \citep{Shi2010,Troxel2012}.
The combined PS and BS can strongly constrain not only the cosmological parameters but also the IA. 

\subsection{Bispectrum covariance}

Thus far, we have not discussed the modeling of BS covariance, which is another important ingredient of cosmological likelihood analysis.
The BS covariance of Gaussian fluctuations has a simple form given by the PS and the shot noise \citep{Sefusatti2006}.
However, in the nonlinear regime, one should consider the non-Gaussian and super-sample contributions \citep[e.g.,][]{TakadaHu2013}, which complicate the evaluation. 
In such cases, the covariance has been estimated by perturbation theory \citep[e.g.,][]{Sugiyama2019}, the halo model \citep[e.g.,][]{Kayo2013,Rizzato2018}, and an ensemble of simulation mocks \citep[e.g.,][]{SN2013,CB2017,CB2018,Cola2019}.
To estimate an unbiased inverse covariance, the last approach should generate more mocks than a number of $k$-bins  \citep[e.g.,][]{Hartlap2007}; consequently, the number of mocks can be huge ($>10^{2\text{--}3}$).
This topic is reserved for future work.

\subsection{Emulator}

Several groups are developing nonlinear PS emulators that interpolate simulation results over a wide range of wavenumbers, redshifts, and cosmological models \citep{Lawrence2017,Garrison2018,Nishimichi2018,Knabe2019,DeRose2019}.
We expect that developing a similar emulator for BS is much more formidable, for two reasons.
First, we measure a {\it binned} BS but require an {\it unbinned} BS (recall that BS is sensitive to binning).
Therefore, we cannot simply interpolate the measured quantities.
Second, BS measurements have larger sample variances than PS measurements, which demand many realizations in each cosmological model.
This is computationally expensive.

%A practical way to develop a BS emulator would be 1) prepare a function of unbinned BS containing several fitting parameters, 2) then fit these parameters for various cosmological models and redshifts, and 3) finally interpolate them.

\section{Conclusions}

\begin{deluxetable*}{ccccc}
\tablecaption{\resp{Calibration range}}
\startdata
\hline 
 Cosmological & minimum $k$ & maximum $k$ & redshift & calibration  \\
 model & ($h \, \Mpc^{-1}$) & ($h \, \Mpc^{-1}$)  &  &  \\
  \hline
 \textit{Planck} 2015 $\Lambda$CDM & $1.6 \times 10^{-3}$  & $14-48$ & $0\text{--}10$ & sim. \& SPT  \\
 \hline
 $40$ $w$CDM & $1.6 \times 10^{-3}$ & $14-28$ & $0\text{--}1.5$ & sim. \& SPT \\
         & $1.6 \times 10^{-3}$ & $0.18-0.28$ & $2\text{--}10$  & SPT
\enddata
\tablecomments{
\resp{Calibration range of $k$ and $z$ in the simulations and the one-loop SPT. The maximum $k$ slightly depends on redshift (see Figure \ref{fig_kmin_kmax}). In the $40$ $w$CDM models at $z=2\text{--}10$, the calibration was done only by SPT. \vspace{0.8cm}}
}
\label{table_fitting_range}
%\vspace{1.0cm}
\end{deluxetable*}

We have constructed a fitting formula of the matter BS calibrated in high-resolution $N$-body simulations of $41$ $w$CDM models around the \textit{Planck} 2015 best-fit $\Lambda$CDM model.
\resp{The calibration covers a wide range of wavenumbers (up to $k= 30 \, h \, \Mpc^{-1}$) and redshifts ($z=0\text{--}10$) for the \textit{Planck} 2015 model. The $40$ $w$CDM models supplement the calibration at $z=0\text{--}1.5$.}
We also performed a large-scale calibration using perturbation theory \resp{for all the cosmological models} (at $k < 0.3 \, h \, \Mpc^{-1}$ and $z=0\text{--}10$).
\resp{The calibration range is summarized in Table~\ref{table_fitting_range}.}
The simulation boxes are sufficiently large (side length $L=1,2,$ and $4 \, h^{-1} \, \Gpc$) to cover almost all triangles ($k_1,k_2,k_3$) measured in forthcoming weak-lensing surveys and CMB lensing experiments. 
The accuracy was within $10 \, (15) \%$ up to $k=3 \, (10) \, h \, \Mpc^{-1}$ in the redshift range $z=0\text{--}3$ for the \textit{Planck} 2015 model.
\resp{The rms deviation was $2.7 \, (3.2) \%$ up to $k=3 \, (10) \, h \, \Mpc^{-1}$ for $z=0\text{--}3$. Therefore, the accuracy was approximately $3 \%$ for most of the triangles and $10\text{--}15 \%$ only for the
worst cases.}
Meanwhile, the accuracy of the $40$ $w$CDM models was around $20 \, \%$ for $k<3 \, h \, \Mpc^{-1}$ and $z=0\text{--}1.5$.
In these models, a $10 \, \%$ intrinsic scatter was introduced to the simulation data by the single realization.
\resp{The rms deviation was $8.0 \, (11.2) \%$ up to $k=3 \, (10) \, h \, \Mpc^{-1}$ for $z=0\text{--}1.5$.}
The user can easily incorporate the baryonic effects (calibrated using IllustrisTNG) into the fitting formula.  
We also confirmed that the formula reproduces the weak-lensing convergence BS measured in light-cone simulations. 

The $\sigma_8$ inferred from the \textit{Planck} results is larger than that estimated from cosmic shear and galaxy-galaxy lensing $\sigma_8$ \cite[$\sigma_8 \sim 0.81$ vs. $\sim 0.77$; e.g.,][]{MacCrann2015,DES2018,Planck2018cosmopara}.
Combining weak-lensing PS and BS can tighten the $\sigma_8$ constraint by a factor of $1.6\text{--}3$ \cite[e.g.,][]{TJ2004,Kayo2013b}, providing new clues for solving this controversy.

\acknowledgements
%We thank ... for their useful comments and discussions. 
We thank the IllustrisTNG team for making their simulations publicly available.
This work was in part supported by Grant-in-Aid for Scientific Research from the Japan Society for the Promotion of Science (JSPS)
(Nos.~JP16H03977, JP17H01131, JP17K14273, JP19H00677, and JP19K14767) and MEXT Grant-in-Aid for Scientific Research on Innovative Areas (No.~JP15H05893, JP15H05896, JP15H05889, JP18H04358, and JP20H04723).
KO is supported by JSPS Overseas Research Fellowships. TN is supported by Japan Science and Technology Agency CREST JPMHCR1414.
Numerical computations were in part carried out on Cray XC30 and XC50 at Centre for Computational Astrophysics, National Astronomical Observatory of Japan.

\appendix

\section{Halo model}

The halo model, which assumes that all matter is confined in halos, is widely applied in nonlinear BS estimation \cite[e.g.,][]{CH2001,CS2002,Valag2011,Kayo2013,Yamamoto2017}. 
The basic properties of a halo of mass $M$ are characterized by the mass function $dn(M)/dM$, the spherical density profile $\rho(r;M)$, and the first- and second-order halo biases $b_{1,2}(M)$.   
This model decomposes the matter BS into three terms: one- (1h), two- (2h), and three-halo (3h) terms.
The 1h and 3h terms dominate at small and large scales, respectively.
The 2h term fills the gap between the 1h and 3h terms and only minimally contributes at intermediate scales, except at the squeezed limit.
%\resp{Therefore, this term is ignored in our fitting formula.}
The BS is given by
\beq
  B(k_1,k_2,k_3) = B^{\rm HM}_{\rm 1h}(k_1,k_2,k_3) \resp{+ B^{\rm HM}_{\rm 2h}(k_1,k_2,k_3)} +  B^{\rm HM}_{\rm 3h}(k_1,k_2,k_3).
\eeq
The 1h term comes from the density profile of a single halo:
\beq
  B^{\rm HM}_{\rm 1h}(k_1,k_2,k_3)=\int \! dM \frac{dn(M)}{dM} \left( \frac{M}{\bar{\rho}} \right)^3 u(k_1;M) u(k_2;M) u(k_3;M) ,
\label{B1h_HM}
\eeq
where $\bar{\rho}$ is the cosmic mean density and $u(k;M)$ is the Fourier transform of the scaled density profile $\rho(r;M)/M$.
\resp{The 2h term describes the correlation among two points in the same halo and a third point in another halo:
\beq
 B^{\rm HM}_{\rm 2h}(k_1,k_2,k_3) = I^1_2(k_1,k_2) I^1_1(k_3) P_{\rm L}(k_3) + 2 \, {\rm perm.},
\label{B2h_HM}
\eeq
with
\beq
    I^1_2(k_1,k_2) = \int \! dM \frac{dn(M)}{dM} \left( \frac{M}{\bar{\rho}} \right)^2 b_1(M) u(k_1;M) u(k_2;M).
\eeq
}
The 3h term describes the spatial correlation among three different halos:
\begin{align}
    B^{\rm HM}_{\rm 3h}(k_1,k_2,k_3) &= I^1_1(k_1) I^1_1(k_2) I^1_1(k_3) B_{\rm tree}(k_1,k_2,k_3) 
    + \left[ I^1_1(k_1) I^1_1(k_2) I^2_1(k_3) P_{\rm L}(k_1) P_{\rm L}(k_2) + 2 \, {\rm perm.} \right] \nonumber \\
    &= 2 \left[ F_2(\bfk_1,\bfk_2) + \frac{I^2_1(k_3)}{2 I^1_1(k_3)} \right] I^1_1(k_1) I^1_1(k_2) I^1_1(k_3) P_{\rm L}(k_1) P_{\rm L}(k_2) + 2 \, {\rm perm.},
\end{align}
with 
\beq
    I^\beta_1(k) = \int \! dM \frac{dn(M)}{dM} \frac{M}{\bar{\rho}} b_\beta(M) u(k;M).
\eeq
\resp{The 2h and 3h terms are proportional to $P_{\rm L}$ and $(P_{\rm L})^2$, respectively.}

\section{Fitting formula}

Our fitting formula adopts the Halofit parameterization for nonlinear PS \citep{Smith2003}.
The dimensionless linear PS is defined as $\Delta_{\rm L}^2(k)=k^3 P_{\rm L}(k)/(2 \pi^2)$.
The nonlinear scale $k_{\rm NL}^{-1}$ is determined as
\beq
  \sigma^2(k_{\rm NL}^{-1})=1, ~~ \sigma^2(R) = \int d \ln k \, \Delta_{\rm L}^2(k) \, {\rm e}^{-k^2 R^2} .
\eeq
The effective spectral index at $k_{\rm NL}$ is defined as
\beq
  n_{\rm eff} + 3 = - \left. \frac{d \ln \sigma^2(R)}{d \ln R} \right|_{R=k_{\rm NL}^{-1}}.
\eeq
We also introduce a scaled wavenumber, $q_{\rm i}=k_{\rm i}/k_{\rm NL}$ (${\rm i}=1,2$ and $3$). 
Note that the quantities $k_{\rm NL}$ and $n_{\rm eff}$ are  evaluated at a given redshift.
Identical parameters were defined in \cite{Smith2003}. 

%We dropped the two-halo term. 
%Instead, we enhance the three-halo term at intermediate scale. 
The fitting function is the sum of the 1h and 3h terms:
\beq
 B(k_1,k_2,k_3) = B_{\rm 1h}(k_1,k_2,k_3) + B_{\rm 3h}(k_1,k_2,k_3).
\label{BS1+3h}
\eeq
The 1h term is
\beq
  B_{\rm 1h}(k_1,k_2,k_3) = \prod_{{\rm i}=1}^3 \left[ \, \frac{1}{a_{\rm n} q_{\rm i}^{\alpha_{\rm n}} + b_{\rm n} q_{\rm i}^{\beta_{\rm n}}} \frac{1}{1+\left( c_{\rm n} q_{\rm i} \right)^{-1}} \right].
\eeq  
Here $B_{\rm 1h}$ is assumed as the product of identical functions of $q_1, q_2,$ and $q_3$. 
Similarly, the halo model $B_{\rm 1h}^{\rm HM}$ given by Eq.~\eqref{B1h_HM} is the product of $u(k_{\rm i})$ terms.
The 3h term is given by
\beq
  B_{\rm 3h}(k_1,k_2,k_3) = 2  \left[ F_2(\bfk_1,\bfk_2) + d_{\rm n} q_3 \right] I(k_1) I(k_2) I(k_3) P_{\rm E}(k_1) P_{\rm E}(k_2) + 2 \, {\rm perm.},
\label{bk_fitting}
\eeq
with
\beq
  P_{\rm E}(k) = \frac{1+f_{\rm n} q^2}{1+ g_{\rm n} q + h_{\rm n} q^2} P_{\rm L}(k) + \frac{1}{m_{\rm n} q^{\mu_{\rm n}} + n_{\rm n} q^{\nu_{\rm n}}} \frac{1}{1+\left( p_{\rm n} q \right)^{-3}}, ~~I(k) = \frac{1}{1+e_{\rm n} q}.
\label{PS_E}
\eeq
Here $P_{\rm E}(k)$ defines the ``enhanced" PS, obtained by adding a small-scale enhancement to the linear PS. 
The first (second) term of $P_{\rm E}$ is similar to the 2h (1h) term of Halofit for the nonlinear PS.
Similarly, $I(k)$ and $d_{\rm n} q$ correspond to $I^1_1(k)$ and $I^2_1(k)/[2 I^1_1(k)]$ in the halo model, respectively.  
This 3h term approaches the tree level in the low-$k$ limit. 

\begin{figure*}
\begin{center}
\includegraphics[width=15cm]{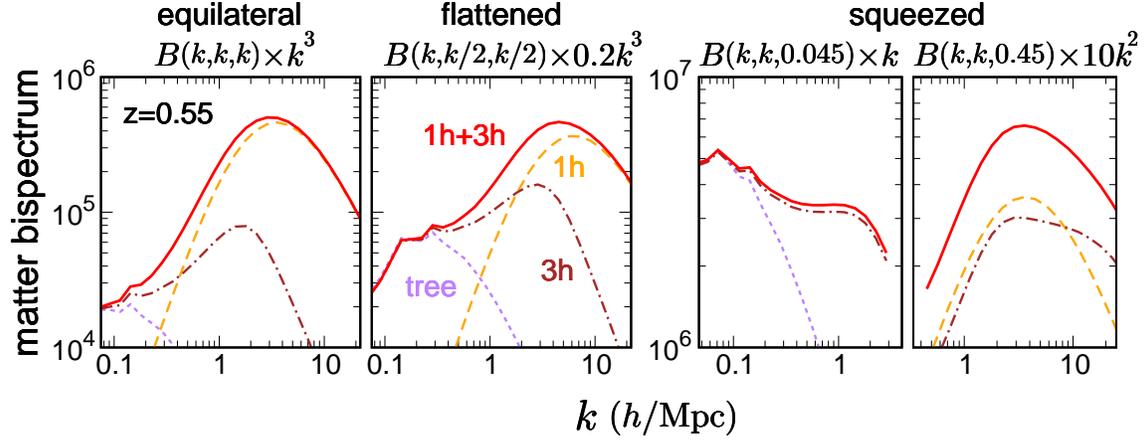}
\end{center}
\caption{Contributions of the one-halo (1h) and three-halo (3h) terms to the total-matter bispectrum in the fitting formula.
\label{fig_bk_1-3haloterms}
}
\end{figure*}

\resp{The 3h term $B_{\rm 3h}$ includes the 2h contribution $B_{\rm 2h}^{\rm HM}$ in the halo model, as discussed below. 
As $B_{\rm 2h}^{\rm HM}$ is proportional to $P_{\rm L}$, several terms proportional to $P_{\rm L}$ in $B_{\rm 3h}$ correspond to $B_{\rm 2h}^{\rm HM}$.
The enhanced PS can be decomposed into the linear PS and the small-scale enhancement: $P_{\rm E}(k) \simeq P_{\rm L}(k)+P_{\rm E}^{\rm 2nd}(k)$ (where the prefactor of $P_{\rm L}$ in Eq.~(\ref{PS_E}) is ignored).
The terms proportional to $P_{\rm L}$ are given from Eq.~(\ref{bk_fitting}) by 
\beq
  \left. B_{\rm 3h}(k_1,k_2,k_3) \right|_{\propto P_{\rm L}} \simeq 2 \left[ \left\{ F_2(\bfk_1,\bfk_3) + d_{\rm n} q_2 \right\} P_{\rm E}^{\rm 2nd}(k_1) + (\bfk_1 \leftrightarrow \bfk_2) \right] I(k_1) I(k_2) I(k_3) P_{\rm L}(k_3) + 2 \, {\rm perm.}  
\label{B3h_propPL}
\eeq
Therefore, $2 [ \{ F_2(\bfk_1,\bfk_3) + d_{\rm n} q_2 \} P_{\rm E}^{\rm 2nd}(k_1) + (\bfk_1 \leftrightarrow \bfk_2) ] I(k_1) I(k_2)$ in the above equation corresponds to $I^1_2(k_1,k_2)$ in 
%the 2h term (\ref{B2h_HM}).
$B_{\rm 2h}^{\rm HM}$. 
Eq.~(\ref{B3h_propPL}) enhances the squeezed $B_{\rm 3h}$ at intermediate scales (see also Figure \ref{fig_bk_1-3haloterms}).
}

The above fitting parameters ($a_{\rm n},b_{\rm n},...$) are polynomials in terms of $n_{\rm eff}$ and $\log_{10} \sigma_8$, where $\sigma_8$ is the spherical overdensity at a radius of $8 \, h^{-1} \, \Mpc$ at redshift $z$ (i.e., $\sigma_8 (z=0)$ multiplied by the linear growth factor).
The amplitude-determining parameters (i.e., $a_{\rm n},b_{\rm n},m_{\rm n}$ and $n_{\rm n}$) are functions of $\log_{10} \sigma_8$, whereas most of the other parameters are functions of $n_{\rm eff}$.
As $B(k_1,k_2,k_3)$ and $P(k)$ have dimensions of $[{\rm L}^6]$ and $[{\rm L}^3]$, respectively, $a_{\rm n}$ and $b_{\rm n}$ have dimensions of $[{\rm L}^{-2}]$, $m_{\rm n}$ and $n_{\rm n}$ have dimensions of $[{\rm L}^{-3}]$, and all other parameters are dimensionless. 
\resp{Here the length unit is chosen as $[{\rm L}]=[h^{-1} \, \Mpc]$.} 
%Though $[{\rm Length}]$ is an arbitrary length unit, one may choose $[h^{-1}\, \Mpc]$ or $[\Mpc]$.  

The fitting parameters of the 1h term are given by
\begin{align}
    \log_{10} a_{\rm n} &= -2.167-2.944 \log_{10} \sigma_8 -1.106 \left( \log_{10} \sigma_8 \right)^2 -2.865 \left( \log_{10} \sigma_8 \right)^3 -0.310 \, r_1^{\gamma_{\rm n}}, \nonumber \\
    \log_{10} b_{\rm n} &= -3.428 -2.681 \log_{10} \sigma_8 +1.624 \left( \log_{10} \sigma_8 \right)^2 -0.095 \left( \log_{10} \sigma_8 \right)^3,  \nonumber \\
    \log_{10} c_{\rm n} &= 0.159 - 1.107 \, n_{\rm eff},  \nonumber \\
    \log_{10} \alpha_{\rm n} &= {\rm min} \left[ -4.348 -3.006 \, n_{\rm eff} -0.5745 \,  n_{\rm eff}^2 + 10^{-0.9+0.2 \, n_{\rm eff}} \, r_2^2, ~\log_{10} \left( 1-\frac{2}{3}n_{\rm s} \right) \right],   \nonumber \\ 
    \log_{10} \beta_{\rm n} &= -1.731 -2.845 \, n_{\rm eff} -1.4995 \, n_{\rm eff}^2 -0.2811 \, n_{\rm eff}^3 +0.007 \, r_2, \nonumber \\
    \log_{10} \gamma_{\rm n} &= 0.182+0.570 \, n_{\rm eff},   
\end{align}
where $r_{1,2}$ are ratios of the minimum ($k_{\rm min}$) and middle ($k_{\rm mid}$) wavenumbers to the maximum ($k_{\rm max}$) wavenumber of the triangle, respectively, given by
\beq
  r_1=\frac{k_{\rm min}}{k_{\rm max}}, ~r_2=\frac{k_{\rm mid}+k_{\rm min}-k_{\rm max}}{k_{\rm max}}.
\eeq
The $r_{1,2}$ represents a ``halo triaxiality'' in the 1h term \citep{Smith2006}: $r_{1,2} \rightarrow 0$ in the squeezed case ($k_{\rm min} \ll k_{\rm mid} \simeq k_{\rm max}$), $r_1 \, (r_2) \rightarrow 0.5 \, (0)$ in the flattened case ($k_{\rm min} \simeq k_{\rm mid} \simeq k_{\rm max}/2$), and $r_{1,2} \rightarrow 1$ in the equilateral case ($k_{\rm min} \simeq k_{\rm mid} \simeq k_{\rm max}$). 
These terms slightly enhance (suppress) the squeezed (equilateral) BS at $k \gtrsim 5 \, h \, \Mpc^{-1}$. 
To ensure that the 1h term is smaller than the tree level in the low-$k$ limit, the maximum $\alpha_{\rm n}$ was set to $\alpha_{\rm n,max}=1-(2/3)n_{\rm s}$ (where $n_{\rm s}$ is the spectral index of the initial PS).

The parameters of the 3h term are given by
\begin{align}
    \log_{10} f_{\rm n} &= -10.533 -16.838 \, n_{\rm eff} -9.3048 \, n_{\rm eff}^2 -1.8263 \, n_{\rm eff}^3,   \nonumber \\
    \log_{10} g_{\rm n} &= 2.787 +2.405 \, n_{\rm eff} +0.4577 \, n_{\rm eff}^2,   \nonumber \\
    \log_{10} h_{\rm n} &= -1.118-0.394 \, n_{\rm eff},  \nonumber \\
    \log_{10} m_{\rm n} &= -2.605 -2.434 \log_{10} \sigma_8 +5.710 \left( \log_{10} \sigma_8 \right)^2,   \nonumber \\
    \log_{10} n_{\rm n} &= -4.468 -3.080 \log_{10} \sigma_8 +1.035 \left( \log_{10} \sigma_8 \right)^2,   \nonumber \\
    \log_{10} \mu_{\rm n} &= 15.312 +22.977 \, n_{\rm eff} +10.9579 \, n_{\rm eff}^2 +1.6586 \, n_{\rm eff}^3,   \nonumber \\ 
    \log_{10} \nu_{\rm n} &= 1.347 +1.246 \, n_{\rm eff} +0.4525 \, n_{\rm eff}^2, \nonumber \\
    \log_{10} p_{\rm n} &= 0.071-0.433 \, n_{\rm eff},  \nonumber \\
    \log_{10} d_{\rm n} &= -0.483 +0.892 \log_{10} \sigma_8 -0.086 \, \Omega_{\rm m}, \nonumber \\
    \log_{10} e_{\rm n} &= -0.632+0.646 \, n_{\rm eff}.
\end{align}
Here $\Omega_{\rm m}$ is the matter density parameter at $z$.
\resp{Note again that $a_{\rm n}$ and $b_{\rm n}$ ($m_{\rm n}$ and $n_{\rm n}$) have the units of $h^2 \, \Mpc^{-2}$ $(h^3 \, \Mpc^{-3})$ and the other parameters are dimensionless.}
As the calibration was performed from $z=0$ to $10$, the formula should be switched to the tree level at $z>10$.

\resp{We checked that the above fitting parameters ($a_{\rm n}, b_{\rm n}, ...$) did not depend on other cosmological parameters as follows: we fitted the formula to the \textit{Planck} 2015 model at each redshift and to each $w$CDM model at $z=0$: then, it turned out that the best-fit values of them ($a_{\rm n}, b_{\rm n},...$) mainly depend on two parameters of $n_{\rm eff}$ and $\sigma_8$, and did not correlate with the other parameters (i.e., cosmological parameters and redshift).}

Figure~\ref{fig_bk_1-3haloterms} plots the separate contributions of $B_{\rm 1h}$ and $B_{\rm 3h}$ at $z=0.55$ in the \textit{Planck} 2015 model.
The results are unbinned.
In the equilateral and flattened cases, the 1h (3h) term clearly dominated at small (large) scales.
In the squeezed case, the 1h (3h) term dominated in the nonlinear (linear) regime of $k_3$. 
\resp{In all cases, the second term of $P_{\rm E}$ enhances the 3h term at intermediate scales ($k \simeq 1\text{--}10 \, h \, \Mpc^{-1}$).}

\begin{figure*}
\begin{center}
\includegraphics[width=15cm]{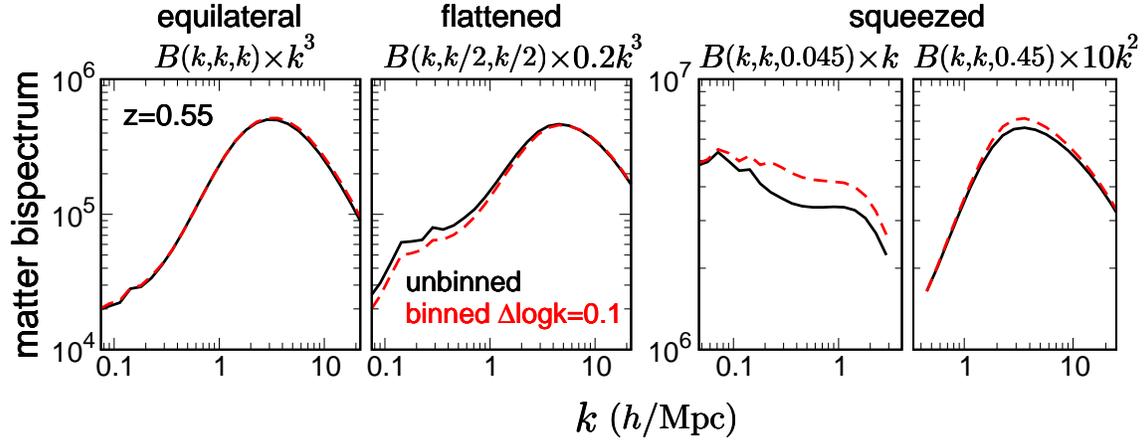}  
\end{center}
\caption{Effect of $k-$binning on the bispectrum fitting formula. The black curve is the unbinned result, and the dashed red curve is the binned result with a bin width of $\Delta \log_{10} k=0.1$.  
\label{fig_bk_1-3haloterms_bin-unbin}
}
\end{figure*}

Figure~\ref{fig_bk_1-3haloterms_bin-unbin} shows the binning effect on BS. The binning affected the squeezed BS, because the cosine term in the $F_2$ kernel is very sensitive to the squeezed triangle configuration \cite[for details, see section~IIB of][]{Namikawa2019}. 

\section{Fitting the ratio of the bispectrum with baryons to that without baryons}

This appendix fits the ratio of the BS with baryons to that without baryons.
The ratio, defined as $R_{\rm b}$ in Eq.~\eqref{ratio_baryon}, was calibrated in the TNG300-1 simulation \citep{Nelson2018}.
The analysis included all triangle configurations $(k_1,k_2,k_3)$ satisfying the following two conditions: a) the number of triangles in the bin exceeds $10^6$ to remove noisy data points, and b) the shot-noise contribution is less than $3 \, \%$.
The fitting range was $k=0.03 \text{--} 30 \, h \, \Mpc^{-1}$ and $z=0\text{--}\resp{10}$ \resp{(eleven redshifts of $z=0, 0.2, 0.4, 0.7, 1, 1.5, 2, 3, 5, 7$ and $10$)}.

\resp{For $z \leq 5$,} the BS ratio $R_{\rm b}$ was given by
\beq
  R_{\rm b}(k_1,k_2,k_3) = \prod_{{\rm i}=1}^{3} \left[ A_0 \exp \left\{ - \left| \frac{x_{\rm i}-\mu_0}{\sigma_0} \right|^{\alpha_0} \right\} - A_1  \exp \left\{ - \left( \frac{x_{\rm i}-\mu_1}{\sigma_1} \right)^2 \right\} + \left\{ \left( \frac{k_{\rm i}}{k_*} \right)^{\alpha_2} +1 \right\}^{\beta_2} \right],
\label{baryon_ratio_fit}
\eeq
where $x_{\rm i}=\log_{\rm 10} [k_{\rm i}/(h \, \Mpc^{-1})]$.
The fitting parameters are the following functions of the scale factor $a$:
\begin{align}
 A_0 &= 0.068 \left( a- 0.5 \right)^{0.47} \Theta(a-0.5), \nonumber \\
 \mu_0 &= 0.018 \, a + 0.837 \, a^2, \nonumber \\
 \sigma_0 &= 0.881 \, \mu_0, \nonumber \\
 \alpha_0 &= 2.346, \nonumber \\
 A_1 &= 1.052 \left( a- 0.2 \right)^{1.41} \Theta(a-0.2), \nonumber \\
 \mu_1 &= \left| 0.172 + 3.048 \, a - 0.675 \, a^2 \right|, \nonumber \\
 \sigma_1 &= (0.494-0.039 \, a) \, \mu_1, \nonumber \\
 k_* &= 29.90 - 38.73 \, a + 24.30 \, a^2, \nonumber \\
 \alpha_2 &= 2.25, \nonumber \\
 \alpha_2 \beta_2 &= \frac{0.563}{\left( {a}/{0.06} \right)^{0.02}+1} ,
\end{align}
where $k_*$ has units of $h \, {\rm Mpc}^{-1}$ and $\Theta (x)$ is the step function: $\Theta (x)=1$ and $0$ for $x\geq0$ and $x<0$, respectively.
The first term of Eq.~\eqref{baryon_ratio_fit} represents the small enhancement at intermediate scales ($k \simeq 1\text{--}10 \, h \, \Mpc^{-1}$) and low $z \, (<1)$, the second term is the depression at $k \approx 10 \, h \,\Mpc^{-1}$, and the last term is the strong enhancement at high $k \, (\gtrsim 10 \, h \, \Mpc^{-1})$.
The ratio $R_{\rm b}$ approaches unity in the low-$k$ limit. 

\resp{At higher redshifts ($z>5$; $z=7$ and $10$ in our data), $R_{\rm b}=1$ was a good approximation in our fitting range ($k<30 \, h \, \Mpc^{-1}$), because the effects of AGN feedback and star formation on BS were suppressed at high $z$.}

\bibliographystyle{aasjournal}
\bibliography{references}

\begin{thebibliography}{}
\expandafter\ifx\csname natexlab\endcsname\relax\def\natexlab#1{#1}\fi
\providecommand{\url}[1]{\href{#1}{#1}}
\providecommand{\dodoi}[1]{doi:~\href{http://doi.org/#1}{\nolinkurl{#1}}}
\providecommand{\doeprint}[1]{\href{http://ascl.net/#1}{\nolinkurl{http://ascl.net/#1}}}
\providecommand{\doarXiv}[1]{\href{https://arxiv.org/abs/#1}{\nolinkurl{https://arxiv.org/abs/#1}}}

\bibitem[{{Abbott} {et~al.}(2018){Abbott}, {Abdalla}, {Alarcon}, {Aleksi{\'c}},
  {Allam}, {Allen}, {Amara}, {Annis}, {Asorey}, \& {Avila}}]{DES2018}
{Abbott}, T.~M.~C., {Abdalla}, F.~B., {Alarcon}, A., {et~al.} 2018, \prd, 98,
  043526, \dodoi{10.1103/PhysRevD.98.043526}

\bibitem[{{Angulo} {et~al.}(2015){Angulo}, {Foreman}, {Schmittfull}, \&
  {Senatore}}]{Angulo2015}
{Angulo}, R.~E., {Foreman}, S., {Schmittfull}, M., \& {Senatore}, L. 2015,
  \jcap, 10, 039, \dodoi{10.1088/1475-7516/2015/10/039}

\bibitem[{{Bartelmann} \& {Schneider}(2001)}]{BS2001}
{Bartelmann}, M., \& {Schneider}, P. 2001, \physrep, 340, 291,
  \dodoi{10.1016/S0370-1573(00)00082-X}

\bibitem[{{Beck} {et~al.}(2018){Beck}, {Fabbian}, \& {Errard}}]{Beck2018}
{Beck}, D., {Fabbian}, G., \& {Errard}, J. 2018, \prd, 98, 043512,
  \dodoi{10.1103/PhysRevD.98.043512}

\bibitem[{{Berg{\'e}} {et~al.}(2010){Berg{\'e}}, {Amara}, \&
  {R{\'e}fr{\'e}gier}}]{Berge2010}
{Berg{\'e}}, J., {Amara}, A., \& {R{\'e}fr{\'e}gier}, A. 2010, The
  Astrophysical Journal, 712, 992, \dodoi{10.1088/0004-637X/712/2/992}

\bibitem[{{Bernardeau} {et~al.}(2002{\natexlab{a}}){Bernardeau}, {Colombi},
  {Gazta{\~n}aga}, \& {Scoccimarro}}]{BCGS2002}
{Bernardeau}, F., {Colombi}, S., {Gazta{\~n}aga}, E., \& {Scoccimarro}, R.
  2002{\natexlab{a}}, \physrep, 367, 1, \dodoi{10.1016/S0370-1573(02)00135-7}

\bibitem[{{Bernardeau} {et~al.}(2002{\natexlab{b}}){Bernardeau}, {Mellier}, \&
  {van Waerbeke}}]{BMvW2002}
{Bernardeau}, F., {Mellier}, Y., \& {van Waerbeke}, L. 2002{\natexlab{b}},
  \aap, 389, L28, \dodoi{10.1051/0004-6361:20020700}

\bibitem[{{B{\"o}hm} {et~al.}(2016){B{\"o}hm}, {Schmittfull}, \&
  {Sherwin}}]{Bohm2016}
{B{\"o}hm}, V., {Schmittfull}, M., \& {Sherwin}, B.~D. 2016, \prd, 94, 043519,
  \dodoi{10.1103/PhysRevD.94.043519}

\bibitem[{{Bose} {et~al.}(2019){Bose}, {Byun}, {Lacasa}, {Moradinezhad Dizgah},
  \& {Lombriser}}]{Bose2019}
{Bose}, B., {Byun}, J., {Lacasa}, F., {Moradinezhad Dizgah}, A., \&
  {Lombriser}, L. 2019, arXiv e-prints, arXiv:1909.02504.
\newblock \doarXiv{1909.02504}

\bibitem[{{Bose} \& {Taruya}(2018)}]{Bose2018}
{Bose}, B., \& {Taruya}, A. 2018, \jcap, 10, 019,
  \dodoi{10.1088/1475-7516/2018/10/019}

\bibitem[{{Byun} {et~al.}(2017){Byun}, {Eggemeier}, {Regan}, {Seery}, \&
  {Smith}}]{Byun2017}
{Byun}, J., {Eggemeier}, A., {Regan}, D., {Seery}, D., \& {Smith}, R.~E. 2017,
  \mnras, 471, 1581, \dodoi{10.1093/mnras/stx1681}

\bibitem[{{Chan} \& {Blot}(2017)}]{CB2017}
{Chan}, K.~C., \& {Blot}, L. 2017, \prd, 96, 023528,
  \dodoi{10.1103/PhysRevD.96.023528}

\bibitem[{{Chan} {et~al.}(2018){Chan}, {Moradinezhad Dizgah}, \&
  {Nore{\~n}a}}]{CB2018}
{Chan}, K.~C., {Moradinezhad Dizgah}, A., \& {Nore{\~n}a}, J. 2018, \prd, 97,
  043532, \dodoi{10.1103/PhysRevD.97.043532}

\bibitem[{{Chang} {et~al.}(2018){Chang}, {Pujol}, {Mawdsley}, {Bacon},
  {Elvin-Poole}, {Melchior}, {Kov{\'a}cs}, {Jain}, {Leistedt}, {Giannantonio},
  {Alarcon}, {Baxter}, {Bechtol}, {Becker}, {Benoit-L{\'e}vy}, {Bernstein},
  {Bonnett}, {Busha}, {Carnero Rosell}, {Castander}, {Cawthon}, {da Costa},
  {Davis}, {De Vicente}, {DeRose}, {Drlica-Wagner}, {Fosalba}, {Gatti},
  {Gaztanaga}, {Gruen}, {Gschwend}, {Hartley}, {Hoyle}, {Huff}, {Jarvis},
  {Jeffrey}, {Kacprzak}, {Lin}, {MacCrann}, {Maia}, {Ogando}, {Prat}, {Rau},
  {Rollins}, {Roodman}, {Rozo}, {Rykoff}, {Samuroff}, {S{\'a}nchez},
  {Sevilla-Noarbe}, {Sheldon}, {Troxel}, {Varga}, {Vielzeuf}, {Vikram},
  {Wechsler}, {Zuntz}, {Abbott}, {Abdalla}, {Allam}, {Annis}, {Bertin},
  {Brooks}, {Buckley-Geer}, {Burke}, {Carrasco Kind}, {Carretero}, {Crocce},
  {Cunha}, {D'Andrea}, {Desai}, {Diehl}, {Dietrich}, {Doel}, {Estrada}, {Fausti
  Neto}, {Fernandez}, {Flaugher}, {Frieman}, {Garc{\'\i}a-Bellido}, {Gruendl},
  {Gutierrez}, {Honscheid}, {James}, {Jeltema}, {Johnson}, {Johnson}, {Kent},
  {Kirk}, {Krause}, {Kuehn}, {Kuhlmann}, {Lahav}, {Li}, {Lima}, {March},
  {Martini}, {Menanteau}, {Miquel}, {Mohr}, {Neilsen}, {Nichol}, {Petravick},
  {Plazas}, {Romer}, {Sako}, {Sanchez}, {Scarpine}, {Schubnell}, {Smith},
  {Smith}, {Soares-Santos}, {Sobreira}, {Suchyta}, {Tarle}, {Thomas}, {Tucker},
  {Walker}, {Wester}, {Zhang}, \& {DES Collaboration}}]{Cahng2018}
{Chang}, C., {Pujol}, A., {Mawdsley}, B., {et~al.} 2018, \mnras, 475, 3165,
  \dodoi{10.1093/mnras/stx3363}

\bibitem[{{Chisari} {et~al.}(2018){Chisari}, {Richardson}, {Devriendt},
  {Dubois}, {Schneider}, {Le Brun}, {Beckmann}, {Peirani}, {Slyz}, \&
  {Pichon}}]{Chisari2018}
{Chisari}, N.~E., {Richardson}, M.~L.~A., {Devriendt}, J., {et~al.} 2018,
  \mnras, 480, 3962, \dodoi{10.1093/mnras/sty2093}

\bibitem[{{Chisari} {et~al.}(2019){Chisari}, {Mead}, {Joudaki}, {Ferreira},
  {Schneider}, {Mohr}, {Tr{\"o}ster}, {Alonso}, {McCarthy}, {Martin-Alvarez},
  {Devriendt}, {Slyz}, \& {van Daalen}}]{Chisari2019}
{Chisari}, N.~E., {Mead}, A.~J., {Joudaki}, S., {et~al.} 2019, The Open Journal
  of Astrophysics, 2, 4, \dodoi{10.21105/astro.1905.06082}

\bibitem[{{Colavincenzo} {et~al.}(2019){Colavincenzo}, {Sefusatti}, {Monaco},
  {Blot}, {Crocce}, {Lippich}, {S{\'a}nchez}, {Alvarez}, {Agrawal}, {Avila},
  {Balaguera-Antol{\'\i}nez}, {Bond}, {Codis}, {Dalla Vecchia}, {Dorta},
  {Fosalba}, {Izard}, {Kitaura}, {Pellejero-Ibanez}, {Stein}, {Vakili}, \&
  {Yepes}}]{Cola2019}
{Colavincenzo}, M., {Sefusatti}, E., {Monaco}, P., {et~al.} 2019, \mnras, 482,
  4883, \dodoi{10.1093/mnras/sty2964}

\bibitem[{{Cooray} \& {Hu}(2001)}]{CH2001}
{Cooray}, A., \& {Hu}, W. 2001, The Astrophysical Journal, 548, 7,
  \dodoi{10.1086/318660}

\bibitem[{{Cooray} \& {Sheth}(2002)}]{CS2002}
{Cooray}, A., \& {Sheth}, R. 2002, \physrep, 372, 1,
  \dodoi{10.1016/S0370-1573(02)00276-4}

\bibitem[{{Coulton} {et~al.}(2019){Coulton}, {Liu}, {Madhavacheril},
  {B{\"o}hm}, \& {Spergel}}]{Coulton2019}
{Coulton}, W.~R., {Liu}, J., {Madhavacheril}, M.~S., {B{\"o}hm}, V., \&
  {Spergel}, D.~N. 2019, Journal of Cosmology and Astro-Particle Physics, 5,
  043, \dodoi{10.1088/1475-7516/2019/05/043}

\bibitem[{{Crocce} {et~al.}(2006){Crocce}, {Pueblas}, \&
  {Scoccimarro}}]{CPS2006}
{Crocce}, M., {Pueblas}, S., \& {Scoccimarro}, R. 2006, \mnras, 373, 369,
  \dodoi{10.1111/j.1365-2966.2006.11040.x}

\bibitem[{{DeRose} {et~al.}(2019){DeRose}, {Wechsler}, {Tinker}, {Becker},
  {Mao}, {McClintock}, {McLaughlin}, {Rozo}, \& {Zhai}}]{DeRose2019}
{DeRose}, J., {Wechsler}, R.~H., {Tinker}, J.~L., {et~al.} 2019, \apj, 875, 69,
  \dodoi{10.3847/1538-4357/ab1085}

\bibitem[{{Fabbian} {et~al.}(2019){Fabbian}, {Lewis}, \& {Beck}}]{Fabbian2019}
{Fabbian}, G., {Lewis}, A., \& {Beck}, D. 2019, arXiv e-prints,
  arXiv:1906.08760.
\newblock \doarXiv{1906.08760}

\bibitem[{{Foreman} {et~al.}(2019){Foreman}, {Coulton}, {Villaescusa-Navarro},
  \& {Barreira}}]{Foreman2019}
{Foreman}, S., {Coulton}, W., {Villaescusa-Navarro}, F., \& {Barreira}, A.
  2019, arXiv e-prints, arXiv:1910.03597.
\newblock \doarXiv{1910.03597}

\bibitem[{{Fry} \& {Gaztanaga}(1993)}]{Fry1993}
{Fry}, J.~N., \& {Gaztanaga}, E. 1993, \apj, 413, 447, \dodoi{10.1086/173015}

\bibitem[{{Fu} {et~al.}(2014){Fu}, {Kilbinger}, {Erben}, {Heymans},
  {Hildebrandt}, {Hoekstra}, {Kitching}, {Mellier}, {Miller}, \&
  {Semboloni}}]{Fu2014}
{Fu}, L., {Kilbinger}, M., {Erben}, T., {et~al.} 2014, \mnras, 441, 2725,
  \dodoi{10.1093/mnras/stu754}

\bibitem[{{Garrison} {et~al.}(2018){Garrison}, {Eisenstein}, {Ferrer},
  {Tinker}, {Pinto}, \& {Weinberg}}]{Garrison2018}
{Garrison}, L.~H., {Eisenstein}, D.~J., {Ferrer}, D., {et~al.} 2018, \apjs,
  236, 43, \dodoi{10.3847/1538-4365/aabfd3}

\bibitem[{{Gatti} {et~al.}(2019){Gatti}, {Chang}, {Friedrich}, {Jain}, {Bacon},
  {Crocce}, {DeRose}, {Ferrero}, {Fosalba}, {Gaztanaga}, {Gruen}, {Harrison},
  {Jeffrey}, {MacCrann}, {McClintock}, {Secco}, {Whiteway}, {Abbott}, {Allam},
  {Annis}, {Avila}, {Brooks}, {Buckley-Geer}, {Burke}, {Carnero Rosell},
  {Carrasco Kind}, {Carretero}, {Cawthon}, {Crocce}, {da Costa}, {De Vicente},
  {Desai}, {Diehl}, {Doel}, {Eifler}, {Estrada}, {Everett}, {Evrard},
  {Fosalba}, {Frieman}, {Garcia-Bellido}, {Gerdes}, {Gruendl}, {Gschwend},
  {Gutierrez}, {James}, {Johnson}, {Krause}, {Kuehn}, {Kuehn}, {Lima}, {Maia},
  {March}, {Marshall}, {Melchior}, {Menanteau}, {Miquel}, {Palmese},
  {Paz-Chinchon}, {Plazas}, {Sanchez}, {Sanchez}, {Scarpine}, {Schubnell},
  {Serrano}, {Sevilla-Noarbe}, {Smith}, {Soares-Santos}, {Suchyta}, {Swanson},
  {Tarle}, {Thomas}, {Troxel}, {Zuntz}, \& {the DES collaboration}}]{Gatti2019}
{Gatti}, M., {Chang}, C., {Friedrich}, O., {et~al.} 2019, arXiv e-prints,
  arXiv:1911.05568.
\newblock \doarXiv{1911.05568}

\bibitem[{{Gil-Mar{\'{\i}}n} {et~al.}(2012){Gil-Mar{\'{\i}}n}, {Wagner},
  {Fragkoudi}, {Jimenez}, \& {Verde}}]{GM12}
{Gil-Mar{\'{\i}}n}, H., {Wagner}, C., {Fragkoudi}, F., {Jimenez}, R., \&
  {Verde}, L. 2012, \jcap, 2, 047, \dodoi{10.1088/1475-7516/2012/02/047}

\bibitem[{{Gil-Mar{\'\i}n} {et~al.}(2016){Gil-Mar{\'\i}n}, {Percival},
  {Cuesta}, {Brownstein}, {Chuang}, {Ho}, {Kitaura}, {Maraston}, {Prada},
  {Rodr{\'\i}guez-Torres}, {Ross}, {Schlegel}, {Schneider}, {Thomas}, {Tinker},
  {Tojeiro}, {Vargas Maga{\~n}a}, \& {Zhao}}]{Gilmarin2016}
{Gil-Mar{\'\i}n}, H., {Percival}, W.~J., {Cuesta}, A.~J., {et~al.} 2016,
  \mnras, 460, 4210, \dodoi{10.1093/mnras/stw1264}

\bibitem[{{G{\'o}rski} {et~al.}(2005){G{\'o}rski}, {Hivon}, {Banday},
  {Wandelt}, {Hansen}, {Reinecke}, \& {Bartelmann}}]{Gorski2005}
{G{\'o}rski}, K.~M., {Hivon}, E., {Banday}, A.~J., {et~al.} 2005, \apj, 622,
  759, \dodoi{10.1086/427976}

\bibitem[{{Groth} \& {Peebles}(1977)}]{Groth1977}
{Groth}, E.~J., \& {Peebles}, P.~J.~E. 1977, \apj, 217, 385,
  \dodoi{10.1086/155588}

\bibitem[{{Hamana} {et~al.}(2019){Hamana}, {Shirasaki}, {Miyazaki}, {Hikage},
  {Oguri}, {More}, {Armstrong}, {Leauthaud}, {Mandelbaum}, \&
  {Miyatake}}]{Hamana2019}
{Hamana}, T., {Shirasaki}, M., {Miyazaki}, S., {et~al.} 2019, arXiv e-prints,
  arXiv:1906.06041.
\newblock \doarXiv{1906.06041}

\bibitem[{Hanson {et~al.}(2009)Hanson, Rocha, \& Gorski}]{Hanson:2009:noise}
Hanson, D., Rocha, G., \& Gorski, K. 2009, \mnras, 400, 2169

\bibitem[{{Harnois-D{\'e}raps} {et~al.}(2015){Harnois-D{\'e}raps}, {van
  Waerbeke}, {Viola}, \& {Heymans}}]{Harnois2015}
{Harnois-D{\'e}raps}, J., {van Waerbeke}, L., {Viola}, M., \& {Heymans}, C.
  2015, \mnras, 450, 1212, \dodoi{10.1093/mnras/stv646}

\bibitem[{{Hartlap} {et~al.}(2007){Hartlap}, {Simon}, \&
  {Schneider}}]{Hartlap2007}
{Hartlap}, J., {Simon}, P., \& {Schneider}, P. 2007, \aap, 464, 399,
  \dodoi{10.1051/0004-6361:20066170}

\bibitem[{{Hashimoto} {et~al.}(2017){Hashimoto}, {Rasera}, \&
  {Taruya}}]{Hashimoto2017}
{Hashimoto}, I., {Rasera}, Y., \& {Taruya}, A. 2017, \prd, 96, 043526,
  \dodoi{10.1103/PhysRevD.96.043526}

\bibitem[{{Hearin} {et~al.}(2012){Hearin}, {Zentner}, \& {Ma}}]{Hearin2012}
{Hearin}, A.~P., {Zentner}, A.~R., \& {Ma}, Z. 2012, \jcap, 4, 034,
  \dodoi{10.1088/1475-7516/2012/04/034}

\bibitem[{{Heitmann} {et~al.}(2009){Heitmann}, {Higdon}, {White}, {Habib},
  {Williams}, {Lawrence}, \& {Wagner}}]{Heitmann2009}
{Heitmann}, K., {Higdon}, D., {White}, M., {et~al.} 2009, \apj, 705, 156,
  \dodoi{10.1088/0004-637X/705/1/156}

\bibitem[{{Hellwing} {et~al.}(2016){Hellwing}, {Schaller}, {Frenk}, {Theuns},
  {Schaye}, {Bower}, \& {Crain}}]{Hellwing2016}
{Hellwing}, W.~A., {Schaller}, M., {Frenk}, C.~S., {et~al.} 2016, \mnras, 461,
  L11, \dodoi{10.1093/mnrasl/slw081}

\bibitem[{{Hikage} {et~al.}(2019){Hikage}, {Oguri}, {Hamana}, {More},
  {Mandelbaum}, {Takada}, {K{\"o}hlinger}, {Miyatake}, {Nishizawa}, \&
  {Aihara}}]{Hikage2019}
{Hikage}, C., {Oguri}, M., {Hamana}, T., {et~al.} 2019, \pasj, 71, 43,
  \dodoi{10.1093/pasj/psz010}

\bibitem[{{Hinshaw} {et~al.}(2013){Hinshaw}, {Larson}, {Komatsu}, {Spergel},
  {Bennett}, {Dunkley}, {Nolta}, {Halpern}, {Hill}, {Odegard}, {Page}, {Smith},
  {Weiland}, {Gold}, {Jarosik}, {Kogut}, {Limon}, {Meyer}, {Tucker}, {Wollack},
  \& {Wright}}]{wmap9yr2013}
{Hinshaw}, G., {Larson}, D., {Komatsu}, E., {et~al.} 2013, \apjs, 208, 19,
  \dodoi{10.1088/0067-0049/208/2/19}

\bibitem[{Hu \& Okamoto(2002)}]{HuOkamoto:2001}
Hu, W., \& Okamoto, T. 2002, \apj, 574, 566

\bibitem[{{Huterer} \& {Takada}(2005)}]{HT2005}
{Huterer}, D., \& {Takada}, M. 2005, Astroparticle Physics, 23, 369,
  \dodoi{10.1016/j.astropartphys.2005.02.006}

\bibitem[{{Jarvis} {et~al.}(2004){Jarvis}, {Bernstein}, \& {Jain}}]{Jarvis2004}
{Jarvis}, M., {Bernstein}, G., \& {Jain}, B. 2004, \mnras, 352, 338,
  \dodoi{10.1111/j.1365-2966.2004.07926.x}

\bibitem[{{Jenkins} {et~al.}(1998){Jenkins}, {Frenk}, {Pearce}, {Thomas},
  {Colberg}, {White}, {Couchman}, {Peacock}, {Efstathiou}, \&
  {Nelson}}]{Jenkins1998}
{Jenkins}, A., {Frenk}, C.~S., {Pearce}, F.~R., {et~al.} 1998, \apj, 499, 20,
  \dodoi{10.1086/305615}

\bibitem[{{Jing}(2005)}]{Jing2005}
{Jing}, Y.~P. 2005, \apj, 620, 559, \dodoi{10.1086/427087}

\bibitem[{{Jing} \& {B{\"o}rner}(1998)}]{Jing1998}
{Jing}, Y.~P., \& {B{\"o}rner}, G. 1998, \apj, 503, 37, \dodoi{10.1086/305997}

\bibitem[{{Joachimi} {et~al.}(2015){Joachimi}, {Cacciato}, {Kitching},
  {Leonard}, {Mandelbaum}, {Sch{\"a}fer}, {Sif{\'o}n}, {Hoekstra}, {Kiessling},
  {Kirk}, \& {Rassat}}]{Joachimi2015}
{Joachimi}, B., {Cacciato}, M., {Kitching}, T.~D., {et~al.} 2015, \ssr, 193, 1,
  \dodoi{10.1007/s11214-015-0177-4}

\bibitem[{{Kayo} \& {Takada}(2013)}]{Kayo2013b}
{Kayo}, I., \& {Takada}, M. 2013, arXiv e-prints.
\newblock \doarXiv{1306.4684}

\bibitem[{{Kayo} {et~al.}(2013){Kayo}, {Takada}, \& {Jain}}]{Kayo2013}
{Kayo}, I., {Takada}, M., \& {Jain}, B. 2013, \mnras, 429, 344,
  \dodoi{10.1093/mnras/sts340}

\bibitem[{{Kayo} {et~al.}(2004){Kayo}, {Suto}, {Nichol}, {Pan}, {Szapudi},
  {Connolly}, {Gardner}, {Jain}, {Kulkarni}, {Matsubara}, {Sheth}, {Szalay}, \&
  {Brinkmann}}]{Kayo2004}
{Kayo}, I., {Suto}, Y., {Nichol}, R.~C., {et~al.} 2004, \pasj, 56, 415,
  \dodoi{10.1093/pasj/56.3.415}

\bibitem[{{Kilbinger} \& {Schneider}(2005)}]{Kilbin2005}
{Kilbinger}, M., \& {Schneider}, P. 2005, \aap, 442, 69,
  \dodoi{10.1051/0004-6361:20053531}

\bibitem[{{Kilbinger} {et~al.}(2017){Kilbinger}, {Heymans}, {Asgari},
  {Joudaki}, {Schneider}, {Simon}, {Van Waerbeke}, {Harnois-D{\'e}raps},
  {Hildebrandt}, {K{\"o}hlinger}, {Kuijken}, \& {Viola}}]{Kilbin2017}
{Kilbinger}, M., {Heymans}, C., {Asgari}, M., {et~al.} 2017, \mnras, 472, 2126,
  \dodoi{10.1093/mnras/stx2082}

\bibitem[{{Kitching} {et~al.}(2017){Kitching}, {Alsing}, {Heavens}, {Jimenez},
  {McEwen}, \& {Verde}}]{Kitching2017}
{Kitching}, T.~D., {Alsing}, J., {Heavens}, A.~F., {et~al.} 2017, \mnras, 469,
  2737, \dodoi{10.1093/mnras/stx1039}

\bibitem[{{Knabenhans} {et~al.}(2019){Knabenhans}, {Stadel}, {Marelli},
  {Potter}, {Teyssier}, {Legrand}, {Schneider}, {Sudret}, {Blot}, {Awan},
  {Burigana}, {Carvalho}, {Kurki-Suonio}, {Sirri}, \& {Euclid
  Collaboration}}]{Knabe2019}
{Knabenhans}, M., {Stadel}, J., {Marelli}, S., {et~al.} 2019, \mnras, 484,
  5509, \dodoi{10.1093/mnras/stz197}

\bibitem[{{Lawrence} {et~al.}(2017){Lawrence}, {Heitmann}, {Kwan}, {Upadhye},
  {Bingham}, {Habib}, {Higdon}, {Pope}, {Finkel}, \&
  {Frontiere}}]{Lawrence2017}
{Lawrence}, E., {Heitmann}, K., {Kwan}, J., {et~al.} 2017, \apj, 847, 50,
  \dodoi{10.3847/1538-4357/aa86a9}

\bibitem[{{Lazanu} {et~al.}(2016){Lazanu}, {Giannantonio}, {Schmittfull}, \&
  {Shellard}}]{Lazanu2016}
{Lazanu}, A., {Giannantonio}, T., {Schmittfull}, M., \& {Shellard}, E.~P.~S.
  2016, Physical Review D, 93, 083517, \dodoi{10.1103/PhysRevD.93.083517}

\bibitem[{{Lazanu} \& {Liguori}(2018)}]{LL2018}
{Lazanu}, A., \& {Liguori}, M. 2018, \jcap, 4, 055,
  \dodoi{10.1088/1475-7516/2018/04/055}

\bibitem[{{Lewis} \& {Challinor}(2006)}]{LC2006}
{Lewis}, A., \& {Challinor}, A. 2006, \physrep, 429, 1,
  \dodoi{10.1016/j.physrep.2006.03.002}

\bibitem[{{Lewis} {et~al.}(2000){Lewis}, {Challinor}, \& {Lasenby}}]{CAMB2000}
{Lewis}, A., {Challinor}, A., \& {Lasenby}, A. 2000, \apj, 538, 473,
  \dodoi{10.1086/309179}

\bibitem[{{Lewis} \& {Pratten}(2016)}]{LP2016}
{Lewis}, A., \& {Pratten}, G. 2016, \jcap, 12, 003,
  \dodoi{10.1088/1475-7516/2016/12/003}

\bibitem[{{MacCrann} {et~al.}(2015){MacCrann}, {Zuntz}, {Bridle}, {Jain}, \&
  {Becker}}]{MacCrann2015}
{MacCrann}, N., {Zuntz}, J., {Bridle}, S., {Jain}, B., \& {Becker}, M.~R. 2015,
  \mnras, 451, 2877, \dodoi{10.1093/mnras/stv1154}

\bibitem[{Madhavacheril \& Hill(2018)}]{Madhavacheril:2018bxi}
Madhavacheril, M.~S., \& Hill, J.~C. 2018, Phys. Rev., D98, 023534,
  \dodoi{10.1103/PhysRevD.98.023534}

\bibitem[{{Marinacci} {et~al.}(2018){Marinacci}, {Vogelsberger}, {Pakmor},
  {Torrey}, {Springel}, {Hernquist}, {Nelson}, {Weinberger}, {Pillepich},
  {Naiman}, \& {Genel}}]{Marinacci2018}
{Marinacci}, F., {Vogelsberger}, M., {Pakmor}, R., {et~al.} 2018, \mnras, 480,
  5113, \dodoi{10.1093/mnras/sty2206}

\bibitem[{{Marozzi} {et~al.}(2018){Marozzi}, {Fanizza}, {Di Dio}, \&
  {Durrer}}]{Marozzi2018}
{Marozzi}, G., {Fanizza}, G., {Di Dio}, E., \& {Durrer}, R. 2018, \prd, 98,
  023535, \dodoi{10.1103/PhysRevD.98.023535}

\bibitem[{{Matarrese} {et~al.}(1997){Matarrese}, {Verde}, \&
  {Heavens}}]{Matar1997}
{Matarrese}, S., {Verde}, L., \& {Heavens}, A.~F. 1997, \mnras, 290, 651,
  \dodoi{10.1093/mnras/290.4.651}

\bibitem[{{McCarthy} {et~al.}(2017){McCarthy}, {Schaye}, {Bird}, \& {Le
  Brun}}]{McCarthy2017}
{McCarthy}, I.~G., {Schaye}, J., {Bird}, S., \& {Le Brun}, A. M.~C. 2017,
  \mnras, 465, 2936, \dodoi{10.1093/mnras/stw2792}

\bibitem[{{McCullagh} {et~al.}(2016){McCullagh}, {Jeong}, \&
  {Szalay}}]{McCullagh2016}
{McCullagh}, N., {Jeong}, D., \& {Szalay}, A.~S. 2016, \mnras, 455, 2945,
  \dodoi{10.1093/mnras/stv2525}

\bibitem[{{Mead} {et~al.}(2015){Mead}, {Peacock}, {Heymans}, {Joudaki}, \&
  {Heavens}}]{Mead2015}
{Mead}, A.~J., {Peacock}, J.~A., {Heymans}, C., {Joudaki}, S., \& {Heavens},
  A.~F. 2015, \mnras, 454, 1958, \dodoi{10.1093/mnras/stv2036}

\bibitem[{Mishra \& Schaan(2019)}]{Mishra:2019qyd}
Mishra, N., \& Schaan, E. 2019.
\newblock \doarXiv{1908.08057}

\bibitem[{{Munshi} {et~al.}(2019){Munshi}, {Namikawa}, {Kitching}, {McEwen},
  {Takahashi}, {Bouchet}, {Taruya}, \& {Bose}}]{Munshi2019}
{Munshi}, D., {Namikawa}, T., {Kitching}, T.~D., {et~al.} 2019, arXiv e-prints,
  arXiv:1910.04627.
\newblock \doarXiv{1910.04627}

\bibitem[{{Munshi} {et~al.}(2011){Munshi}, {Smidt}, {Heavens}, {Coles}, \&
  {Cooray}}]{Munshi2011}
{Munshi}, D., {Smidt}, J., {Heavens}, A., {Coles}, P., \& {Cooray}, A. 2011,
  \mnras, 411, 2241, \dodoi{10.1111/j.1365-2966.2010.17838.x}

\bibitem[{{Naiman} {et~al.}(2018){Naiman}, {Pillepich}, {Springel},
  {Ramirez-Ruiz}, {Torrey}, {Vogelsberger}, {Pakmor}, {Nelson}, {Marinacci},
  {Hernquist}, {Weinberger}, \& {Genel}}]{Naiman2018}
{Naiman}, J.~P., {Pillepich}, A., {Springel}, V., {et~al.} 2018, \mnras, 477,
  1206, \dodoi{10.1093/mnras/sty618}

\bibitem[{{Namikawa}(2016)}]{Namikawa2016}
{Namikawa}, T. 2016, \prd, 93, 121301, \dodoi{10.1103/PhysRevD.93.121301}

\bibitem[{{Namikawa} {et~al.}(2019){Namikawa}, {Bose}, {Bouchet}, {Takahashi},
  \& {Taruya}}]{Namikawa2019}
{Namikawa}, T., {Bose}, B., {Bouchet}, F.~R., {Takahashi}, R., \& {Taruya}, A.
  2019, \prd, 99, 063511, \dodoi{10.1103/PhysRevD.99.063511}

\bibitem[{Namikawa {et~al.}(2013)Namikawa, Hanson, \&
  Takahashi}]{Namikawa:2012:bias-hardening}
Namikawa, T., Hanson, D., \& Takahashi, R. 2013, \mnras, 431, 609,
  \dodoi{10.1093/mnras/stt195}

\bibitem[{{Nelson} {et~al.}(2018){Nelson}, {Pillepich}, {Springel},
  {Weinberger}, {Hernquist}, {Pakmor}, {Genel}, {Torrey}, {Vogelsberger},
  {Kauffmann}, {Marinacci}, \& {Naiman}}]{Nelson2018a}
{Nelson}, D., {Pillepich}, A., {Springel}, V., {et~al.} 2018, \mnras, 475, 624,
  \dodoi{10.1093/mnras/stx3040}

\bibitem[{{Nelson} {et~al.}(2019){Nelson}, {Springel}, {Pillepich},
  {Rodriguez-Gomez}, {Torrey}, {Genel}, {Vogelsberger}, {Pakmor}, {Marinacci},
  {Weinberger}, {Kelley}, {Lovell}, {Diemer}, \& {Hernquist}}]{Nelson2018}
{Nelson}, D., {Springel}, V., {Pillepich}, A., {et~al.} 2019, Computational
  Astrophysics and Cosmology, 6, 2, \dodoi{10.1186/s40668-019-0028-x}

\bibitem[{{Nishimichi} {et~al.}(2007){Nishimichi}, {Kayo}, {Hikage}, {Yahata},
  {Taruya}, {Jing}, {Sheth}, \& {Suto}}]{Nishimichi2007}
{Nishimichi}, T., {Kayo}, I., {Hikage}, C., {et~al.} 2007, \pasj, 59, 93,
  \dodoi{10.1093/pasj/59.1.93}

\bibitem[{{Nishimichi} {et~al.}(2009){Nishimichi}, {Shirata}, {Taruya},
  {Yahata}, {Saito}, {Suto}, {Takahashi}, {Yoshida}, {Matsubara}, {Sugiyama},
  {Kayo}, {Jing}, \& {Yoshikawa}}]{Nishimichi2009}
{Nishimichi}, T., {Shirata}, A., {Taruya}, A., {et~al.} 2009, \pasj, 61, 321,
  \dodoi{10.1093/pasj/61.2.321}

\bibitem[{{Nishimichi} {et~al.}(2019){Nishimichi}, {Takada}, {Takahashi},
  {Osato}, {Shirasaki}, {Oogi}, {Miyatake}, {Oguri}, {Murata}, {Kobayashi}, \&
  {Yoshida}}]{Nishimichi2018}
{Nishimichi}, T., {Takada}, M., {Takahashi}, R., {et~al.} 2019, \apj, 884, 29,
  \dodoi{10.3847/1538-4357/ab3719}

\bibitem[{{Osato} {et~al.}(2015){Osato}, {Shirasaki}, \& {Yoshida}}]{Osato2015}
{Osato}, K., {Shirasaki}, M., \& {Yoshida}, N. 2015, \apj, 806, 186,
  \dodoi{10.1088/0004-637X/806/2/186}

\bibitem[{Osborne {et~al.}(2014)Osborne, Hanson, \& Dore}]{Osborne:2013nna}
Osborne, S.~J., Hanson, D., \& Dore, O. 2014, \jcap, 03, 024

\bibitem[{{Peebles} \& {Groth}(1975)}]{Peebles1975}
{Peebles}, P.~J.~E., \& {Groth}, E.~J. 1975, \apj, 196, 1,
  \dodoi{10.1086/153390}

\bibitem[{{Petri} {et~al.}(2015){Petri}, {Liu}, {Haiman}, {May}, {Hui}, \&
  {Kratochvil}}]{Petri2015}
{Petri}, A., {Liu}, J., {Haiman}, Z., {et~al.} 2015, \prd, 91, 103511,
  \dodoi{10.1103/PhysRevD.91.103511}

\bibitem[{{Pillepich} {et~al.}(2018){Pillepich}, {Nelson}, {Hernquist},
  {Springel}, {Pakmor}, {Torrey}, {Weinberger}, {Genel}, {Naiman}, {Marinacci},
  \& {Vogelsberger}}]{Pillep2018}
{Pillepich}, A., {Nelson}, D., {Hernquist}, L., {et~al.} 2018, \mnras, 475,
  648, \dodoi{10.1093/mnras/stx3112}

\bibitem[{{Planck Collaboration}(2016)}]{Planck2015}
{Planck Collaboration}. 2016, \aap, 594, A13,
  \dodoi{10.1051/0004-6361/201525830}

\bibitem[{{Planck Collaboration}(2018{\natexlab{a}})}]{Planck2018lens}
---. 2018{\natexlab{a}}, arXiv e-prints, arXiv:1807.06210.
\newblock \doarXiv{1807.06210}

\bibitem[{{Planck Collaboration}(2018{\natexlab{b}})}]{Planck2018cosmopara}
---. 2018{\natexlab{b}}, arXiv e-prints, arXiv:1807.06209.
\newblock \doarXiv{1807.06209}

\bibitem[{{Planck Collaboration}(2019)}]{Planck2018nonG}
---. 2019, arXiv e-prints, arXiv:1905.05697.
\newblock \doarXiv{1905.05697}

\bibitem[{{Pratten} \& {Lewis}(2016)}]{PL2016}
{Pratten}, G., \& {Lewis}, A. 2016, \jcap, 8, 047,
  \dodoi{10.1088/1475-7516/2016/08/047}

\bibitem[{{Press} {et~al.}(2002){Press}, {Teukolsky}, {Vetterling}, \&
  {Flannery}}]{NumericalRecipes}
{Press}, W.~H., {Teukolsky}, S.~A., {Vetterling}, W.~T., \& {Flannery}, B.~P.
  2002, {Numerical recipes in C++ : the art of scientific computing}

\bibitem[{{Rampf} \& {Wong}(2012)}]{RW2012}
{Rampf}, C., \& {Wong}, Y. Y.~Y. 2012, \jcap, 6, 018,
  \dodoi{10.1088/1475-7516/2012/06/018}

\bibitem[{{Rizzato} {et~al.}(2018){Rizzato}, {Benabed}, {Bernardeau}, \&
  {Lacasa}}]{Rizzato2018}
{Rizzato}, M., {Benabed}, K., {Bernardeau}, F., \& {Lacasa}, F. 2018, arXiv
  e-prints, arXiv:1812.07437.
\newblock \doarXiv{1812.07437}

\bibitem[{{Sato} {et~al.}(2009){Sato}, {Hamana}, {Takahashi}, {Takada},
  {Yoshida}, {Matsubara}, \& {Sugiyama}}]{Sato2009}
{Sato}, M., {Hamana}, T., {Takahashi}, R., {et~al.} 2009, \apj, 701, 945,
  \dodoi{10.1088/0004-637X/701/2/945}

\bibitem[{{Sato} \& {Nishimichi}(2013)}]{SN2013}
{Sato}, M., \& {Nishimichi}, T. 2013, \prd, 87, 123538,
  \dodoi{10.1103/PhysRevD.87.123538}

\bibitem[{{Schaye} {et~al.}(2015){Schaye}, {Crain}, {Bower}, {Furlong},
  {Schaller}, {Theuns}, {Dalla Vecchia}, {Frenk}, {McCarthy}, {Helly},
  {Jenkins}, {Rosas-Guevara}, {White}, {Baes}, {Booth}, {Camps}, {Navarro},
  {Qu}, {Rahmati}, {Sawala}, {Thomas}, \& {Trayford}}]{Schaye2015}
{Schaye}, J., {Crain}, R.~A., {Bower}, R.~G., {et~al.} 2015, \mnras, 446, 521,
  \dodoi{10.1093/mnras/stu2058}

\bibitem[{{Schneider} {et~al.}(2016){Schneider}, {Teyssier}, {Potter},
  {Stadel}, {Onions}, {Reed}, {Smith}, {Springel}, {Pearce}, \&
  {Scoccimarro}}]{Schneider2016}
{Schneider}, A., {Teyssier}, R., {Potter}, D., {et~al.} 2016, \jcap, 4, 047,
  \dodoi{10.1088/1475-7516/2016/04/047}

\bibitem[{{Scoccimarro}(1997)}]{Scocci1997}
{Scoccimarro}, R. 1997, \apj, 487, 1, \dodoi{10.1086/304578}

\bibitem[{{Scoccimarro}(2015)}]{Scocci2015}
---. 2015, \prd, 92, 083532, \dodoi{10.1103/PhysRevD.92.083532}

\bibitem[{{Scoccimarro} {et~al.}(1998){Scoccimarro}, {Colombi}, {Fry},
  {Frieman}, {Hivon}, \& {Melott}}]{Scocci1998}
{Scoccimarro}, R., {Colombi}, S., {Fry}, J.~N., {et~al.} 1998, \apj, 496, 586,
  \dodoi{10.1086/305399}

\bibitem[{{Scoccimarro} \& {Couchman}(2001)}]{SC01}
{Scoccimarro}, R., \& {Couchman}, H.~M.~P. 2001, \mnras, 325, 1312,
  \dodoi{10.1046/j.1365-8711.2001.04281.x}

\bibitem[{{Scoccimarro} {et~al.}(2001){Scoccimarro}, {Feldman}, {Fry}, \&
  {Frieman}}]{Scocci2001}
{Scoccimarro}, R., {Feldman}, H.~A., {Fry}, J.~N., \& {Frieman}, J.~A. 2001,
  \apj, 546, 652, \dodoi{10.1086/318284}

\bibitem[{{Scoccimarro} \& {Frieman}(1999)}]{SF1999}
{Scoccimarro}, R., \& {Frieman}, J.~A. 1999, \apj, 520, 35,
  \dodoi{10.1086/307448}

\bibitem[{{Sefusatti} {et~al.}(2010){Sefusatti}, {Crocce}, \&
  {Desjacques}}]{Sefusatti2010}
{Sefusatti}, E., {Crocce}, M., \& {Desjacques}, V. 2010, \mnras, 406, 1014,
  \dodoi{10.1111/j.1365-2966.2010.16723.x}

\bibitem[{{Sefusatti} {et~al.}(2006){Sefusatti}, {Crocce}, {Pueblas}, \&
  {Scoccimarro}}]{Sefusatti2006}
{Sefusatti}, E., {Crocce}, M., {Pueblas}, S., \& {Scoccimarro}, R. 2006, \prd,
  74, 023522, \dodoi{10.1103/PhysRevD.74.023522}

\bibitem[{{Sefusatti} {et~al.}(2016){Sefusatti}, {Crocce}, {Scoccimarro}, \&
  {Couchman}}]{Sefusatti2016}
{Sefusatti}, E., {Crocce}, M., {Scoccimarro}, R., \& {Couchman}, H.~M.~P. 2016,
  \mnras, 460, 3624, \dodoi{10.1093/mnras/stw1229}

\bibitem[{{Semboloni} {et~al.}(2008){Semboloni}, {Heymans}, {van Waerbeke}, \&
  {Schneider}}]{Sembo2008}
{Semboloni}, E., {Heymans}, C., {van Waerbeke}, L., \& {Schneider}, P. 2008,
  \mnras, 388, 991, \dodoi{10.1111/j.1365-2966.2008.13478.x}

\bibitem[{{Semboloni} {et~al.}(2013){Semboloni}, {Hoekstra}, \&
  {Schaye}}]{Sembo2013}
{Semboloni}, E., {Hoekstra}, H., \& {Schaye}, J. 2013, \mnras, 434, 148,
  \dodoi{10.1093/mnras/stt1013}

\bibitem[{{Semboloni} {et~al.}(2011{\natexlab{a}}){Semboloni}, {Hoekstra},
  {Schaye}, {van Daalen}, \& {McCarthy}}]{Sembo2011b}
{Semboloni}, E., {Hoekstra}, H., {Schaye}, J., {van Daalen}, M.~P., \&
  {McCarthy}, I.~G. 2011{\natexlab{a}}, \mnras, 417, 2020,
  \dodoi{10.1111/j.1365-2966.2011.19385.x}

\bibitem[{{Semboloni} {et~al.}(2011{\natexlab{b}}){Semboloni}, {Schrabback},
  {van Waerbeke}, {Vafaei}, {Hartlap}, \& {Hilbert}}]{Sembo2011}
{Semboloni}, E., {Schrabback}, T., {van Waerbeke}, L., {et~al.}
  2011{\natexlab{b}}, \mnras, 410, 143,
  \dodoi{10.1111/j.1365-2966.2010.17430.x}

\bibitem[{{Shi} {et~al.}(2010){Shi}, {Joachimi}, \& {Schneider}}]{Shi2010}
{Shi}, X., {Joachimi}, B., \& {Schneider}, P. 2010, \aap, 523, A60,
  \dodoi{10.1051/0004-6361/201014191}

\bibitem[{{Shirasaki} {et~al.}(2015){Shirasaki}, {Hamana}, \&
  {Yoshida}}]{Shirasaki2015}
{Shirasaki}, M., {Hamana}, T., \& {Yoshida}, N. 2015, \mnras, 453, 3043,
  \dodoi{10.1093/mnras/stv1854}

\bibitem[{{Simon} {et~al.}(2015){Simon}, {Semboloni}, {van Waerbeke},
  {Hoekstra}, {Erben}, {Fu}, {Harnois-D{\'e}raps}, {Heymans}, {Hildebrandt}, \&
  {Kilbinger}}]{Simon2015}
{Simon}, P., {Semboloni}, E., {van Waerbeke}, L., {et~al.} 2015, \mnras, 449,
  1505, \dodoi{10.1093/mnras/stv339}

\bibitem[{{Slepian} {et~al.}(2017){Slepian}, {Eisenstein}, {Brownstein},
  {Chuang}, {Gil-Mar{\'\i}n}, {Ho}, {Kitaura}, {Percival}, {Ross}, {Rossi},
  {Seo}, {Slosar}, \& {Vargas-Maga{\~n}a}}]{Slepian2017}
{Slepian}, Z., {Eisenstein}, D.~J., {Brownstein}, J.~R., {et~al.} 2017, \mnras,
  469, 1738, \dodoi{10.1093/mnras/stx488}

\bibitem[{{Smith} \& {Angulo}(2019)}]{SA2019}
{Smith}, R.~E., \& {Angulo}, R.~E. 2019, \mnras, 486, 1448,
  \dodoi{10.1093/mnras/stz890}

\bibitem[{{Smith} {et~al.}(2006){Smith}, {Watts}, \& {Sheth}}]{Smith2006}
{Smith}, R.~E., {Watts}, P.~I.~R., \& {Sheth}, R.~K. 2006, \mnras, 365, 214,
  \dodoi{10.1111/j.1365-2966.2005.09707.x}

\bibitem[{{Smith} {et~al.}(2003){Smith}, {Peacock}, {Jenkins}, {White},
  {Frenk}, {Pearce}, {Thomas}, {Efstathiou}, \& {Couchman}}]{Smith2003}
{Smith}, R.~E., {Peacock}, J.~A., {Jenkins}, A., {et~al.} 2003, \mnras, 341,
  1311, \dodoi{10.1046/j.1365-8711.2003.06503.x}

\bibitem[{{Spergel} {et~al.}(2007){Spergel}, {Bean}, {Dor{\'e}}, {Nolta},
  {Bennett}, {Dunkley}, {Hinshaw}, {Jarosik}, {Komatsu}, {Page}, {Peiris},
  {Verde}, {Halpern}, {Hill}, {Kogut}, {Limon}, {Meyer}, {Odegard}, {Tucker},
  {Weiland}, {Wollack}, \& {Wright}}]{Spergel2007}
{Spergel}, D.~N., {Bean}, R., {Dor{\'e}}, O., {et~al.} 2007, \apjs, 170, 377,
  \dodoi{10.1086/513700}

\bibitem[{{Springel}(2005)}]{Springel2005}
{Springel}, V. 2005, \mnras, 364, 1105,
  \dodoi{10.1111/j.1365-2966.2005.09655.x}

\bibitem[{{Springel} {et~al.}(2001){Springel}, {Yoshida}, \&
  {White}}]{Springel2001}
{Springel}, V., {Yoshida}, N., \& {White}, S.~D.~M. 2001, New Astron., 6, 79,
  \dodoi{10.1016/S1384-1076(01)00042-2}

\bibitem[{{Springel} {et~al.}(2018){Springel}, {Pakmor}, {Pillepich},
  {Weinberger}, {Nelson}, {Hernquist}, {Vogelsberger}, {Genel}, {Torrey},
  {Marinacci}, \& {Naiman}}]{Springel2018}
{Springel}, V., {Pakmor}, R., {Pillepich}, A., {et~al.} 2018, \mnras, 475, 676,
  \dodoi{10.1093/mnras/stx3304}

\bibitem[{{Sugiyama} {et~al.}(2019){Sugiyama}, {Saito}, {Beutler}, \&
  {Seo}}]{Sugiyama2019}
{Sugiyama}, N.~S., {Saito}, S., {Beutler}, F., \& {Seo}, H.-J. 2019, arXiv
  e-prints, arXiv:1908.06234.
\newblock \doarXiv{1908.06234}

\bibitem[{{Takada} \& {Hu}(2013)}]{TakadaHu2013}
{Takada}, M., \& {Hu}, W. 2013, \prd, 87, 123504,
  \dodoi{10.1103/PhysRevD.87.123504}

\bibitem[{{Takada} \& {Jain}(2002)}]{TJ2002}
{Takada}, M., \& {Jain}, B. 2002, \mnras, 337, 875,
  \dodoi{10.1046/j.1365-8711.2002.05972.x}

\bibitem[{{Takada} \& {Jain}(2004)}]{TJ2004}
---. 2004, \mnras, 348, 897, \dodoi{10.1111/j.1365-2966.2004.07410.x}

\bibitem[{{Takahashi} {et~al.}(2017){Takahashi}, {Hamana}, {Shirasaki},
  {Namikawa}, {Nishimichi}, {Osato}, \& {Shiroyama}}]{Takahashi2017}
{Takahashi}, R., {Hamana}, T., {Shirasaki}, M., {et~al.} 2017, \apj, 850, 24,
  \dodoi{10.3847/1538-4357/aa943d}

\bibitem[{{Takahashi} {et~al.}(2012){Takahashi}, {Sato}, {Nishimichi},
  {Taruya}, \& {Oguri}}]{Takahasi2012}
{Takahashi}, R., {Sato}, M., {Nishimichi}, T., {Taruya}, A., \& {Oguri}, M.
  2012, \apj, 761, 152, \dodoi{10.1088/0004-637X/761/2/152}

\bibitem[{{Troxel} \& {Ishak}(2012)}]{Troxel2012}
{Troxel}, M.~A., \& {Ishak}, M. 2012, \mnras, 419, 1804,
  \dodoi{10.1111/j.1365-2966.2011.20205.x}

\bibitem[{{Troxel} \& {Ishak}(2015)}]{Troxel2015}
---. 2015, \physrep, 558, 1, \dodoi{10.1016/j.physrep.2014.11.001}

\bibitem[{{Troxel} {et~al.}(2018){Troxel}, {MacCrann}, {Zuntz}, {Eifler},
  {Krause}, {Dodelson}, {Gruen}, {Blazek}, {Friedrich}, {Samuroff}, {Prat},
  {Secco}, {Davis}, {Fert{\'e}}, {DeRose}, {Alarcon}, {Amara}, {Baxter},
  {Becker}, {Bernstein}, {Bridle}, {Cawthon}, {Chang}, {Choi}, {De Vicente},
  {Drlica-Wagner}, {Elvin-Poole}, {Frieman}, {Gatti}, {Hartley}, {Honscheid},
  {Hoyle}, {Huff}, {Huterer}, {Jain}, {Jarvis}, {Kacprzak}, {Kirk}, {Kokron},
  {Krawiec}, {Lahav}, {Liddle}, {Peacock}, {Rau}, {Refregier}, {Rollins},
  {Rozo}, {Rykoff}, {S{\'a}nchez}, {Sevilla-Noarbe}, {Sheldon}, {Stebbins},
  {Varga}, {Vielzeuf}, {Wang}, {Wechsler}, {Yanny}, {Abbott}, {Abdalla},
  {Allam}, {Annis}, {Bechtol}, {Benoit-L{\'e}vy}, {Bertin}, {Brooks},
  {Buckley-Geer}, {Burke}, {Carnero Rosell}, {Carrasco Kind}, {Carretero},
  {Castander}, {Crocce}, {Cunha}, {D'Andrea}, {da Costa}, {DePoy}, {Desai},
  {Diehl}, {Dietrich}, {Doel}, {Fernandez}, {Flaugher}, {Fosalba},
  {Garc{\'\i}a-Bellido}, {Gaztanaga}, {Gerdes}, {Giannantonio}, {Goldstein},
  {Gruendl}, {Gschwend}, {Gutierrez}, {James}, {Jeltema}, {Johnson}, {Johnson},
  {Kent}, {Kuehn}, {Kuhlmann}, {Kuropatkin}, {Li}, {Lima}, {Lin}, {Maia},
  {March}, {Marshall}, {Martini}, {Melchior}, {Menanteau}, {Miquel}, {Mohr},
  {Neilsen}, {Nichol}, {Nord}, {Petravick}, {Plazas}, {Romer}, {Roodman},
  {Sako}, {Sanchez}, {Scarpine}, {Schindler}, {Schubnell}, {Smith}, {Smith},
  {Soares-Santos}, {Sobreira}, {Suchyta}, {Swanson}, {Tarle}, {Thomas},
  {Tucker}, {Vikram}, {Walker}, {Weller}, {Zhang}, \& {DES
  Collaboration}}]{Troxel2018}
{Troxel}, M.~A., {MacCrann}, N., {Zuntz}, J., {et~al.} 2018, \prd, 98, 043528,
  \dodoi{10.1103/PhysRevD.98.043528}

\bibitem[{{Valageas} \& {Nishimichi}(2011)}]{Valag2011}
{Valageas}, P., \& {Nishimichi}, T. 2011, \aap, 532, A4,
  \dodoi{10.1051/0004-6361/201116638}

\bibitem[{{Valageas} {et~al.}(2012){Valageas}, {Sato}, \&
  {Nishimichi}}]{Valag2012}
{Valageas}, P., {Sato}, M., \& {Nishimichi}, T. 2012, \aap, 541, A161,
  \dodoi{10.1051/0004-6361/201118549}

\bibitem[{{van Daalen} {et~al.}(2011){van Daalen}, {Schaye}, {Booth}, \& {Dalla
  Vecchia}}]{vanD2011}
{van Daalen}, M.~P., {Schaye}, J., {Booth}, C.~M., \& {Dalla Vecchia}, C. 2011,
  \mnras, 415, 3649, \dodoi{10.1111/j.1365-2966.2011.18981.x}

\bibitem[{van Engelen {et~al.}(2014)van Engelen, Bhattacharya, Sehgal, Holder,
  Zahn, \& Nagai}]{vanEngelen:2013rla}
van Engelen, A., Bhattacharya, S., Sehgal, N., {et~al.} 2014, \apj, 786, 14

\bibitem[{{van Uitert} {et~al.}(2018){van Uitert}, {Joachimi}, {Joudaki},
  {Amon}, {Heymans}, {K{\"o}hlinger}, {Asgari}, {Blake}, {Choi}, \&
  {Erben}}]{KiDS2018}
{van Uitert}, E., {Joachimi}, B., {Joudaki}, S., {et~al.} 2018, \mnras, 476,
  4662, \dodoi{10.1093/mnras/sty551}

\bibitem[{{Van Waerbeke} {et~al.}(2013){Van Waerbeke}, {Benjamin}, {Erben},
  {Heymans}, {Hildebrandt}, {Hoekstra}, {Kitching}, {Mellier}, {Miller}, \&
  {Coupon}}]{Van2013}
{Van Waerbeke}, L., {Benjamin}, J., {Erben}, T., {et~al.} 2013, \mnras, 433,
  3373, \dodoi{10.1093/mnras/stt971}

\bibitem[{{Vogelsberger} {et~al.}(2014){Vogelsberger}, {Genel}, {Springel},
  {Torrey}, {Sijacki}, {Xu}, {Snyder}, {Nelson}, \& {Hernquist}}]{Vogel2014}
{Vogelsberger}, M., {Genel}, S., {Springel}, V., {et~al.} 2014, \mnras, 444,
  1518, \dodoi{10.1093/mnras/stu1536}

\bibitem[{{Yamamoto} {et~al.}(2017){Yamamoto}, {Nan}, \&
  {Hikage}}]{Yamamoto2017}
{Yamamoto}, K., {Nan}, Y., \& {Hikage}, C. 2017, \prd, 95, 043528,
  \dodoi{10.1103/PhysRevD.95.043528}

\end{thebibliography}

\end{document}